\documentclass{article}
\usepackage[english]{babel}
\usepackage[utf8]{inputenc}
\usepackage{johd}

\usepackage{amsmath,amsfonts,amssymb}
\usepackage{graphicx}
\usepackage{epstopdf, epsfig}
\usepackage{xcolor}
\usepackage{float}

\usepackage{bbold}
\usepackage{bbm}
\usepackage{float}
\usepackage{caption}
\usepackage{comment}
\usepackage{booktabs}
\usepackage{verbatim}

\title{Mean- and unsteady-flow reconstruction with one or two time-resolved measurements}

\author{Lucas Franceschini$^{a}$, Denis Sipp$^{a}$, Olivier Marquet$^{a}$ \\
        \small $^{a}$ Department of Aerodynamics, Aeroacoustics and Aeroelasticity, ONERA, Meudon, France\\
}

\providecommand\xx{\mathbf{x}}

\providecommand\bnabla{\mathbf{\nabla}}

\providecommand\uu{\mathbf{u}}
\providecommand\ouu{\overline{\mathbf{u}}}

\providecommand\tuu{\tilde{\mathbf{u}}}

\providecommand\op{\overline{p}}

\providecommand\ff{\mathbf{f}}
\providecommand\off{\overline{\mathbf{f}}}
\providecommand\of{\overline{f}}

\begin{document}

\maketitle

\begin{abstract} 
In this article, we propose a methodology to reconstruct, in a single step, the mean- and unsteady properties of a flow from very few time-resolved measurements. The procedure is based on the {\it a priori} alignement of Fourier- and Resolvent-modes over energetic frequencies, which is a common feature in shear-dominated transitional flows. Hence, the Reynolds-stresses, which determine the mean-flow, may be approximated from a series of Resolvent modes, which discretize the fluctuation field in the frequency space and whose amplitudes can be tuned thanks to few measurements.
In practice, we solve a nonlinear optimization problem (with only few parameters) based on a model coupling strongly the equations governing the mean-flow and the Resolvent modes. The input data for the assimilation procedure may be very sparse, typically one or two pointwise measurements. This technique is applied to two distinct physical configurations, one "oscillator" flow with periodic fluctuations (squared-section cylinder) and one "noise amplifier" flow with a broadband frequency spectrum (backward-facing step).
\end{abstract}

\noindent\\

\section{Introduction}

\noindent The reconstruction of a time-averaged (mean) flow and its fluctuations is a topical issue within data-assimilation in fluid mechanics. Mean-flow reconstruction alone can be performed through a variational minimization algorithm, where tuning parameters of a mean-flow model are adjusted to match available measurement data from experiments or higher-fidelity numerical simulations. \cite{Foures14} applied this technique on the incompressible laminar flow over a circular cylinder. They used a spatially-distributed volume forcing term (modeling the divergence of the Reynolds-stress tensor) as control parameter to reconstruct the laminar mean-flow using mean-velocity measurements. The same technique was employed by \cite{Symon17} on an experimental incompressible flow around an idealized airfoil at much higher Reynolds number. One drawback of this approach is the need for a large number of well-placed measurements for the reconstruction to produce accurate results. This issue was partially addressed by \cite{Foures14} who studied the ability of the algorithm to interpolate/extrapolate the field in regions of the physical domain far from the measurements. In the case of high-Reynolds number turbulent flows, it becomes mandatory to consider a turbulence model. Such a strategy was initially introduced for the field-inversion step to learn new turbulence models \citep{duraisamy2019turbulence}:
for example, \cite{Singh16,Parish16,he2019data,ferrero2020field} considered a spatially distributed correction term acting on the turbulence production in the Spalart-Allmaras model. \cite{franceschini2020mean} actually showed that it may be better to consider a force correction term in the momentum equations when dense measurements are available and a scalar source term in the turbulence equation in the case of sparse measurements.In the case of dense measurements and a force correction term, \citet{Foures14,Symon17,franceschini2020mean} showed that the procedure may lead to an accurate reconstruction of the Reynolds stress tensor (from the force term).

Those techniques remain purely steady approaches, where mean-flow information is used to reconstruct the mean-flow field. If we target the reconstruction of the fluctuating field in the case of few pointwise measurements, we need to resort to unsteady data-assimilation techniques. Classically, those methods belong to two (possibly overlapping) categories: 4DVar \citep{LeDimet86} and Ensemble Kalman filters \citep{Evensen09}. Those methods have been extensively applied in the context of meteorology \citep{Lorenc86,Liu08} and more recently in fluid mechanics \citep{Mons16,Gronskis13,DAdamo07,meldi2018augmented}. Such techniques are computationally intensive since they are based on an unsteady model that is marched forward in time, while the optimization step is performed either by a time-adjoint code (for 4DVar) and/or a statistical treatment of an ensemble of points in the phase space (for Ensemble Kalman filter).

An alternative approach for the reconstruction of the fluctuation field relies on the analysis of the linear Navier-Stokes operator around the mean-flow. The origin of this approach lies in the works of \citet{pier2002frequency,Barkley06}, who observed that, in the case of supercritical cylinder flow, the stability analysis of the linearized Navier-Stokes operator on the time-averaged (mean) flow (instead of the steady solution or the base-flow) gives rise to a mode whose eigenvalue is purely imaginary and its frequency close to the one of the nonlinear limit cycle. The theoretical explanation of this fact was further explored by \cite{Sipp07} with the aid of weakly-nonlinear analysis close to the bifurcation and then by \citet{mezic2013analysis,Turton15}, who showed that the Fourier mode of a harmonic (in contrast to time-periodic) flow corresponds to a marginal eigenmode of the Jacobian operator. Later, this led \cite{ManticLugo15} to develop a self-consistent model describing the nonlinear saturation of such oscillator flow, where the steady solution evolves toward the mean-flow through the action of the Reynolds-stresses, which are approximated from the marginal eigenmode associated to the mean-flow. The extension of the mean-flow analysis to flows characterized by a broadband spectral frequency content has been conducted by \citet{hwang2010amplification,McKeon10} in streamwise homogeneous flows and by \citet{beneddine2016conditions} in more general configurations. They showed that, under some fairly general assumptions (in shear dominated flows, strong spatial convective instabilities such as the Kelvin-Helmholtz instability favour rank-1 Resolvent operators), the Fourier mode of the nonlinear fluctuation field at a given frequency is proportional to the leading mode of the Singular Value Decomposition of the Resolvent operator around the mean-flow, evaluated at that frequency. In stochastic flows, \cite{towne2018spectral} actually showed that the Resolvent modes are equal to the Spectral Proper Orthogonal Decomposition (SPOD) modes in the case of a white noise spatially uncorrelated forcing.
Yet, this is generally not the case in turbulent flows and the alignement between Resolvent and SPOD modes may be less perfect.
The use of an eddy-viscosity field modeling the small-scale turbulence in the Resolvent analysis may strongly increase the alignement of these modes \citep{morra2019relevance}: \citet{pickering2020optimal} showed that it is possible to use and adjoint-based optimization strategy to determine a single eddy-viscosity field that simultaneously improves the alignement of several Resolvent modes at a given frequency in highly turbulent compressible jet flows. 

Coming back to the estimation problem, if the Fourier mode and Resolvent modes are assumed to be aligned at a given frequency, we know the spatial structure of the fluctuation field at that frequency, up to a (complex) multiplicative constant which represents the energy and phase of that mode. \cite{beneddine2016conditions,gomez2016reduced,beneddine2017unsteady,towne2020resolvent} showed how to estimate this constant with few point-wise time-resolved measurements. From a practical viewpoint, the main drawback of this approach is of course the need for the \textit{a priori} knowledge of the mean-flow. \citet{he2019data,symon2019tale,symon2020mean} therefore proposed to combine in serial a variational mean-flow data-assimilation to determine an approximate mean-flow and a Resolvent analysis for the estimation of the unsteady fluctuation field.

In the present work, we propose to combine in a single step both the estimation of the mean-flow and the estimation (by Resolvent analysis) of the fluctuation field. By doing so, we obtain a nonlinear model whose unknown parameters are the amplitudes of the Resolvent modes at each considered frequency. To tune these parameters, we only need few sensors (of the same order as the number of unknowns). For this reason, one could consider this model as a reduced order model for the unsteady flow, which is then used for data-assimilation. Only very sparse information is required here for the reconstruction of the mean- and fluctuating quantities. This is a strong improvement with respect to the uncoupled serial approach, since an accurate recovery of the mean-flow requires denser measurements (at least lines of measurements). This procedure will be applied to simple 2D model problems of transitional laminar flows, leaving therefore aside the need of an eddy-viscosity and the issue of multiple strong SPOD modes: we assume as in \cite{beneddine2017unsteady} that the alignment between dominant Resolvent modes and Fourier modes (if only one SPOD mode is energetic at a given frequency, we will call it a Fourier mode) is high for all energetic frequencies). The first configuration will be the flow around a square cylinder ($Re=100$), falling in the category of oscillator flows, for which the frequency spectrum exhibits a dominant peak and smaller harmonics. This means that only one Resolvent mode may be sufficient at the fundamental frequency (and thus only one parameter) for the reconstruction procedure. We will show that the velocity at some location of the flow then suffices for the estimation of the mean-flow and the fluctuation field. 
The second configuration is the two-dimensional backward-facing step ($Re=500$), a typical example of noise amplifier \cite{Herve12}, where unsteadiness is triggered by upstream located white-noise in time forcing. Here again, the frequency content of the flow at some points can be used to tune a series of amplitudes of Resolvent modes, which discretize the broadband fluctuation field in the frequency space.

The article proceeds as follows. First (\S \ref{sec:theory}), we present the model combining the mean-flow and the perturbation equations. In particular, we show how to solve for this problem for a given set of Resolvent amplitudes. Then, we successively present the results of the data-assimilation procedure for the cylinder flow case (\ref{sec:cylinder}) and then for the backward facing step flow (\ref{sec:BFS}).


\section{Theory} \label{sec:theory}

This section is devoted to the derivation of the model that will be used in the data-assimilation procedure. We consider an incompressible flow, whose velocity and pressure fields are governed by the Navier-Stokes equations written in compact form:
\begin{equation}\label{eqn:NS}
    \partial_t \uu + N(\uu) = \mathbf{0}, \;\;\;\; \text{where} \;\;\;\; N(\uu) = ( \uu \cdot \bnabla ) \uu + \bnabla p - Re^{-1} \Delta \uu
\end{equation}
is the nonlinear Navier-Stokes operator, with the pressure $p$ such that the velocity field is incompressible $\bnabla \cdot \uu = 0$. The Reynolds number $\textit{Re}=U_{\infty} L / \nu $ is based on an arbitrary velocity $U_{\infty}$ and a characteristic length $L$, which are used to make all the variables non-dimensional. Since we focus on statistically steady regimes, we may decompose the instantaneous state into a mean and a fluctuating component, as
\begin{equation}\label{eqn:ReynoldsDecomposition}
	\uu(\xx,t) = \ouu(\xx) + \uu'(\xx,t), \;\;\;\; p(\xx,t) = \op(\xx) + p'(\xx,t),
\end{equation}
where the time-average operator is defined as 
\begin{equation}
    \overline{(\cdot)} = \lim_{T \rightarrow + \infty} \frac{1}{T} \int_0^T (\cdot)(t) dt.
\end{equation}
Introducing the decomposition (\ref{eqn:ReynoldsDecomposition}) into the Navier-Stokes equations (\ref{eqn:NS}) and taking the time-average yields the mean flow equation:
\begin{equation}\label{eqn:MeanNS}
    N(\ouu) = - \bnabla \cdot ( \overline{\uu' \otimes \uu'} ) \equiv \off,
\end{equation}
where the mean-pressure $\op$ is obtained such that $\nabla \cdot \ouu = 0$. The equation defining the mean-flow differs from the fixed point one ($N(\uu)=0$) by the presence of the right-hand side term $\off \equiv - \bnabla \cdot (\overline{\uu' \otimes \uu'})$. The goal here is to model this term by approximating the fluctuation $\mathbf{u}'$ by singular vectors of the Resolvent operator. To do so, we subtract the mean flow equation (\ref{eqn:MeanNS}) from the instantaneous equations (\ref{eqn:NS}), leading to:
\begin{equation}\label{eqn:FluctNS}
    \partial_t \uu' + L_{\ouu} \uu' = \bnabla \cdot ( \overline{\uu' \otimes \uu'} - \uu' \otimes \uu' ) \equiv \ff',
\end{equation}
where the operator $L_{\ouu} \uu' = \ouu \cdot \bnabla \uu' + \uu' \cdot \bnabla \ouu + \bnabla p' - Re^{-1} \Delta \uu'$ is the linearized Navier-Stokes operator and $\ff'$ represents the nonlinear fluctuations, acting as a forcing term. We then try to find solutions of this equation in the frequency domain by using the ansatz $\uu' = \hat{\uu} e^{\mathrm{i} \omega t} + c.c.$ and $\ff' = \hat{\ff} e^{\mathrm{i} \omega t} + c.c.$, leading to:
\begin{equation}\label{eqn:Fourier}
    \hat{\uu}(\omega) = R(\ouu,\omega) \hat{\ff} (\omega), \;\;\;\; \text{where} \;\;\;\; R(\ouu,\omega) \equiv ( \mathrm{i} \omega I + L_{\ouu} )^{-1}
\end{equation}
is the so-called Resolvent operator. 
This equation represents an input/output relation between the forcing term $\hat{\ff}$ and its perturbation $\hat{\uu}$. In order to isolate the most energetic dynamics (\citet{beneddine2016conditions}), we maximize the gains:
\begin{equation}
	G(\omega) = \frac{||\hat{\uu} ||^2}{|| \hat{\ff} ||^2} = \frac{(R^{\dagger}(\ouu;\omega) R(\ouu;\omega) \hat{\ff} , \hat{\ff} )} {( \hat{\ff} , \hat{\ff} )}
\end{equation}
where $R^{\dagger}(\ouu,\omega)$ is the adjoint Resolvent operator, given by the general definition $(R^{\dagger} \mathbf{a},\mathbf{b}) = (\mathbf{a},R \mathbf{b} ), \forall \mathbf{a},\mathbf{b}$. Here we consider the energy inner product $(\mathbf{a},\mathbf{b}) = \int_{\Omega} \mathbf{a}^* \cdot \mathbf{b} \; d\xx$. 
This optimization problem may be solved by considering either of the following eigen-problems:
\begin{equation}\label{eqn:SVD0}
    R^{\dagger}(\overline{\mathbf{u}},\omega) R(\overline{\mathbf{u}};\omega) \hat{\mathbf{f}}_k = \mu^2_k (\omega) \hat{\mathbf{f}}_k, \;\;\;\; R(\overline{\mathbf{u}},\omega) R^{\dagger}(\overline{\mathbf{u}};\omega) \hat{\mathbf{u}}_k = \mu^2_k (\omega) \hat{\mathbf{u}}_k,
\end{equation}
where $\mu^2_k$ are the eigenvalues (ordered such that $\mu^2_1 \geq \mu^2_2 \geq \cdots \geq 0$), $\hat{\mathbf{u}}_k$ and $\hat{\mathbf{f}}_k$ are the optimal responses and forcings, both normalized such that $|| \hat{\mathbf{f}}_k || = || \hat{\mathbf{u}}_k || = 1$. These vectors consist in two orthonormal bases with respect to the chosen inner-product and may be related through $\hat{\mathbf{u}}_k = \mu_k^{-1} R(\overline{\mathbf{u}}; \omega) \hat{\mathbf{f}}_k$.
A Fourier mode of the flowfield $\hat{\mathbf{u}}$ can therefore be linked to the nonlinear forcing mode $\hat{\mathbf{f}}$ through:
\begin{equation}\label{eqn:SVD2a}
	\hat{\mathbf{u}} (\mathbf{x},\omega) = \sum_{k=1}^{+ \infty} \mu_k(\omega) \beta_k (\omega) \hat{\mathbf{u}}_{k}(\mathbf{x},\omega),
\end{equation}
where $\beta_k(\omega) = \left( \hat{\mathbf{f}}_{k}(\omega) , \hat{\mathbf{f}}(\omega) \right)$ is the projection of the nonlinear forcing term onto the right singular-vector.
In shear-dominated flows, it is common that rank-1 approximations of the Resolvent operator may be sufficient to represent the input-output dynamics of the flow, leading to the following approximation of the fluctuation field:
\begin{equation}\label{eqn:SVD2}
	\hat{\mathbf{u}} (\mathbf{x},\omega) \approx \mu_1(\omega) \beta_1 (\omega) \hat{\mathbf{u}}_{1}(\mathbf{x},\omega),
\end{equation}
 At this point, the notations $ \hat{\mathbf{u}}(\omega)$ and $ \hat{\mathbf{f}}(\omega)$ refer to a given spectral decomposition of the flowfield: in the following, we will make a distinction between time-periodic (usually oscillators, for which $\beta_k(\omega)$ is a Dirac-like distribution) and broadband-frequency (usually noise-amplifiers, for which $\beta_k(\omega)$ is broadband) flows.
\subsection{Time-periodic flows}

Many transitional flows present a periodic behaviour such as the flows around bluff bodies. In those cases the steady-state solution is unstable and the instability converges toward a saturated periodic limit-cycle. Furthermore, it is not uncommon that most of the energy of this periodic limit-cycle is concentrated at the fundamental frequency $\omega_0$ (see \cite{Turton15}). For those flows, the harmonic averaging process:
\begin{equation}
    \check{\uu}(\omega_0) = \lim_{T\rightarrow+\infty} \frac{1}{T} \mathcal{F}_T(\uu')(\omega_0), \;\;\;\; \text{where} \;\;\;\; \mathcal{F}_T(\uu')(\omega) = \int_{0}^T e^{- \mathrm{i} \omega t} \uu'(t) dt
\end{equation}
converges for some discrete frequencies $ \omega_0$ and the lowest value corresponds to the fundamental mode of the Fourier series expansion (or, more generally, a Koopman mode of the flow, see \cite{arbabi2017study}).
\citet{mezic2013analysis} and \cite{Turton15} showed that a Koopman mode was close to the marginal eigensolution of the linearized Navier-Stokes operator at the frequency $ \omega_0 $ if the flow is harmonic (see \citet{Barkley02,Sipp07,ManticLugo15}). Here we will approximate the Koopman mode with the Resolvent mode $ \check{\uu}(\omega_0) \approx A \hat{\uu}_1(\omega_0)$, which is justified by the fact that these structures coincide when the eigenvalue is almost purely imaginary. Furthermore, even though the gain $\mu_1(\omega_0)$ is large, the nonlinear forcing coefficient ($\beta_1(\omega_0)$) is generally weak (interaction between the weak second harmonic at $ 2\omega_0 $ and the fundamental at $ \omega_0$), yielding a finite amplitude $A=\mu_1(\omega_0) \beta_1(\omega_0)$. 
The fluctuation then reads:
\begin{equation}\label{eqn:fluctuationpeaked}
	\mathbf{u}' \approx A \hat{\mathbf{u}}_1 e^{\mathrm{i} \omega_0 t} + c.c.,
\end{equation}
where the parameter $A$ is now a single complex scalar. Based on this, we can approximate the Reynolds-stress tensor as $\of=- 2 |A|^2 \Re \{ \hat{\uu}_1 \otimes \hat{\uu}_1^* \}$. Considering the mean flow equation \eqref{eqn:MeanNS}, an approximation $\tilde{\mathbf{u}}$ of the mean-flow $\overline{\mathbf{u}}$ may therefore be obtained through:
\begin{subequations}\label{eqn:Model1}
	\begin{align}
    N(\tuu) & = - |A|^2 \phi( \hat{\uu}_1,\hat{\uu}_1^* ) \\
	R(\tuu,\omega_0) R^{\dagger}(\tuu,\omega_0) \hat{\mathbf{u}}_1 & = \mu^2_1 \hat{\mathbf{u}}_1,
	\end{align}
\end{subequations}
where $^*$ stands for the complex-conjugation and $\phi(\uu_1,\uu_2) = (\uu_1 \cdot \bnabla ) \uu_2 + (\uu_2 \cdot \bnabla ) \uu_1$ is the symmetrized convective operator. We also remind that the dominant optimal response has been normalized such that $ || \hat{\mathbf{u}}_1 ||=1$.
At this point, we remark that the system formed by the mean flow equation and the Resolvent analysis \eqref{eqn:Model1} is a closed set of nonlinear equations, if one knows the frequency of the nonlinear signal $\omega_0$ and its energy $|A|^2$ (or amplitude $|A|$). Arguably, the frequency of the flow may be determined from a time-resolved signal of the flow, for instance a hot-wire or a point-wise wall-pressure measurement. Furthermore, the amplitude of that signal (or any other measure of the flowfield) could be used to determine the amplitude of the resolvent mode $|A|$ since, for every value of $|A|$, one can solve for the nonlinear solution of \eqref{eqn:Model1} and therefore tune this parameter (if the relation is monotonous) so that the solution matches the actual measure. The output of this procedure is the full flow-field $ \mathbf{u} $ composed of the reconstructed mean-flow $ \mathbf{\tilde{u}} $ and the fluctuation $ \mathbf{{u}'} $ given in equation \eqref{eqn:fluctuationpeaked}. The strength of the approach lies in the simplicity of the optimization problem, since it only involves a single parameter contrarily to the procedure described in \cite{Foures14}, where large-scale spatial fields had to be determined. 

\subsection{Broadband Flows}

For broadband flows, the previous approximation is no longer suited. Indeed, since the frequency spectrum is broadband, we need to "discretize" the fluctuation field in the frequency domain. Furthermore, the fluctuation field $\uu'$ does not hold any  Koopman mode (or, alternatively, Fourier-series representation), since the quantity $\check{\uu}(\omega)$ converges to zero for all frequencies $\omega$. For this reason, we consider the statistics of the unsteady flow, contained in the two-point ($\xx$ and $\xx'$), two-time (the configuration being homogeneous in time, a single time $ t $ is required) correlation tensor:
\begin{equation}\label{eqn:crosscorr}
    \mathbf{T}_{\xx,\xx'}(t) := \mathbb{E}[\uu'(\xx,\tau) \uu'(\xx',\tau+t)] = \lim_{T\rightarrow+\infty} \frac{1}{T} \int_0^T \uu'(\xx,\tau) \uu'(\xx',\tau+t) d\tau.
\end{equation}
Due to the absence of Koopman modes, this tensor tends to zero when $t\rightarrow \pm \infty$ and is therefore square-integrable. For this reason, we may consider the Fourier-transform of this quantity to obtain the two-point spectral correlation tensor: \begin{equation}\label{eqn:spectralcorr}
    \hat{\mathbf{T}}_{\xx,\xx'}(\omega) := \int_{-\infty}^{+\infty} e^{- \mathrm{i} \omega t} \mathbf{T}_{\xx,\xx'} (t) \; dt = \lim_{T\rightarrow+\infty} \frac{1}{T} \mathbb{E}[\mathcal{F}_T (\uu'(\xx))(\omega) \mathcal{F}_T (\uu'(\xx'))(\omega)^*]
\end{equation}
If $\xx=\xx'$, $\mathbf{T}_{\xx,\xx}(t)$ is called the auto-correlation function such that, for $t=0$, $\mathbf{T}_{\xx,\xx}(0)$ contains the Reynolds-stress tensor information at $\xx$. $\hat{\mathbf{T}}_{\xx,\xx}(\omega)$ is the power-spectral density (PSD), which contains the distribution of the energy of the flow per frequency at $ \xx $. Using the approximation by the Resolvent analysis \eqref{eqn:SVD2}, this tensor becomes:
\begin{equation}\label{eqn:spectralcorrapprox}
    \hat{\mathbf{T}}_{\xx,\xx'}(\omega) \approx \mu_1^2 \mathbb{E}[|\beta_1|^2] \hat{\uu}_1(\xx,\omega) \hat{\uu}_1^*(\xx',\omega),
\end{equation}
where, since the SVD decompositon is deterministic (since it is based on the mean-flow quantity), the expectation operator acts only on the expansion coefficients $\beta_k(\omega)$. Coming back now to the cross-correlation tensor, applying the inverse Fourier Transform on this tensor, we have:
\begin{equation}\label{eqn:crosscorrapprox}
    \mathbf{T}_{\xx,\xx'}(t) = \frac{1}{2\pi} \int_{-\infty}^{+\infty} \hat{\mathbf{T}}_{\xx,\xx'} (\omega) e^{\mathrm{i} \omega t} d \omega \approx \frac{1}{2\pi} \int_{0}^{+\infty} \mu_1^2 \mathbb{E}[|\beta_1|^2] \hat{\uu}_1(\xx,\omega) \hat{\uu}_1^*(\xx',\omega) d\omega + c.c.
\end{equation}
This approximation is still not adequate for our purpose since it is not yet discrete in frequency. To do so, we consider a "grid" in the frequency space, $\{ \Omega_j \}_{j=1,N+1}$, so that most of the energy of the fluctuation falls in the interval $(\Omega_0,\Omega_{N+1})$. Furthermore, in the following, we set $\Omega_0 = 0$. A further approximation of \eqref{eqn:crosscorrapprox} is finally given by:
\begin{equation}\label{eqn:crosscorrapprox2}
    \mathbf{T}_{\xx,\xx'}(t) \approx \sum_{j=1}^{N} |A_j|^2 \hat{\uu}_1(\xx,\omega_j) \hat{\uu}_1^*(\xx',\omega_j) + c.c.,
\end{equation}
where we conveniently define $|A_j|^2 = (2\pi)^{-1} \mu_1^2(\omega_j) (\Omega_{j+1}-\Omega_{j}) \mathbb{E}[|\beta_1 (\omega_j)|^2]$ and the frequencies $\omega_j$ are defined to be the center point of each of the intervals in frequency $\omega_j = (\Omega_j+\Omega_{j+1})/2$. This approximation leads to a fluctuation $\uu'$ of the following form:
\begin{equation}\label{eqn:fluctuationbroad}
	\mathbf{u}' = \sum_{j = 1}^N A_j \hat{\mathbf{u}}_1(\omega_j) e^{\mathrm{i} \omega_j t} + c.c.
\end{equation}
which is nothing but an extension of the approximation employed for the time-periodic flow to consider multiple frequencies. The approximation of the mean-flow $\tuu$ can be given now using equation \eqref{eqn:MeanNS}, implying that:
\begin{subequations}\label{eqn:Model2}
	\begin{align}
	N(\tuu) & = - \sum_{j=1}^N |A_j|^2 \phi( \hat{\uu}_1(\omega_j),\hat{\uu}_1^*(\omega_j) ) \\
	R(\ouu,\omega_j) R^{\dagger}(\ouu,\omega_j) \hat{\mathbf{u}}_1(\omega_j) & = \mu^2_1 (\omega_j) \hat{\mathbf{u}}_1(\omega_j),
	\end{align}
\end{subequations}
where again all dominant optimal responses are unit-norm.
\begin{figure}
	\centering
    \raisebox{0.7in}{(a) }\includegraphics[trim={0cm 0cm 0cm 0cm},clip,width=120px]{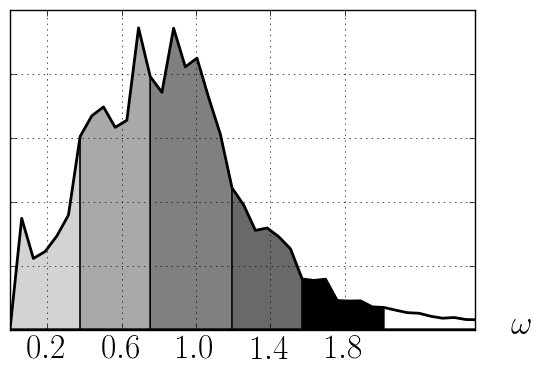}
	\raisebox{0.7in}{(b) }\includegraphics[trim={0cm 0cm 0cm 0cm},clip,width=120px]{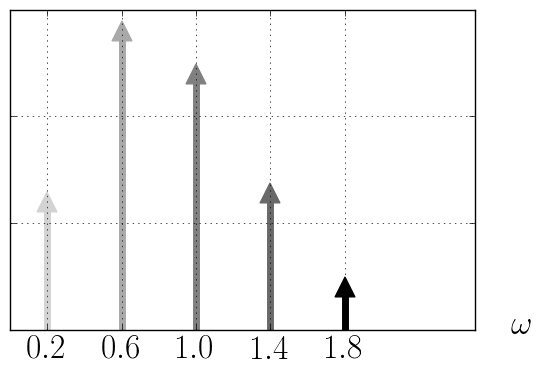}
	\caption{Typical Power Spectral Density (PSD) $\hat{\mathbf{T}}_{\xx,\xx}(\omega)$ for a broadband flow (a). Its integral on five ($N=5$) frequency intervals (represented as the shades of gray) are compared with the discrete-in-frequency energies coming from the model (b).}\label{fig:peaked-spread-signals-white-noise}
\end{figure}

Similarly as before, this system is closed if we provide the amplitudes $|A_j|$ of the Resolvent modes. Again, we will suppose that we have access to (few) point-wise time-resolved probes in the flow, located at some probing points $\{ \xx_p^m \}_{m=1, \cdots, M}$, allowing us to have a good idea not only of the amplitude $|A_j|$ (and its phase) but also of the frequencies $\omega_j$. We need then to establish a comparison basis between the ``discrete-frequency" model \ref{eqn:Model2} and its ``continuous-frequency" counterpart, given by the real flow. To do so, we integrate equation \eqref{eqn:spectralcorrapprox}, evaluated at the measurement points $\xx_p^m = \xx = \xx'$, leading to:
\begin{equation}
    I_{\xx_p^m,\omega_j} \equiv \int_{\Omega_j}^{\Omega_{j+1}} \hat{\mathbf{T}}_{\xx_p^m,\xx_p^m} (\omega) d\omega \approx 2 \pi |A_j|^2 |\hat{\uu}_j(\xx_p^m,\omega_j)|^2.
\end{equation}

In Figure \ref{fig:peaked-spread-signals-white-noise}, we provide a schematic view of this integral: the PSD is divided in a few intervals (resulting now in an energy quantity) and ``lumped" at the discrete set of frequencies $\{ \omega_j \}$.

In terms of practical resolution of the system, the inputs are the quantities $I_{\xx_p^m,\omega_j}$ and the scanning of the full control space $\{ |A_j| \} \in \mathbb{R}^N$ is no longer feasible as soon as $N$ becomes large. For this reason, we will adopt an optimization approach where we will penalise the differences between $I_{\xx_p^m,\omega_j}$ and $2 \pi |A_j|^2 |\hat{\uu}_j(\xx_p^m,\omega_j)|^2$ for some measurement points and frequency bands. We remark that, in theory, only one point ($M=1$) is sufficient for determining all the amplitudes. However, in practise, that point could have weak energies for certain range of frequencies, leading often to poor predictions (as discussed in \citet{beneddine2016conditions}). For this reason, we will choose specific frequency bands within the list of available measurement points. More details on the construction of the cost functional will be given later. Once this cost-functional is chosen, its minimization will be carried out by a gradient-free method, here the COBYLA algorithm (see \citet{conn1997convergence,powell2007view}), from python/scipy's library ``minimize". It is a simplex-based algorithm, which is appropriate if the number of modes $N$ remains reasonably small. For larger $N$ (or if sub-optimal response modes need to be considered) a gradient is needed, which can be obtained from an adjoint-based approach.

\subsubsection{Spectral analysis of broadband flows}

At this point, it is convenient to present the tools to perform a spectral analysis of broadband flows.
The resulting quantities will be compared with the mean-flow Resolvent modes, as done later to assess the efficiency of the fluctuation reconstruction. This analysis is based on the decomposition of the spectral cross-correlation tensor $\hat{\mathbf{T}}_{\xx,\xx'}(\omega)$, which may hold several spatio-temporal coherent modes that can be extracted, from the most energetic to the least. To do so, we employ the Spectral Proper-Orthogonal Decomposition (SPOD, see \citet{towne2018spectral}):
\begin{equation}\label{eqn:SPOD}
    \int_{\Omega} \hat{\mathbf{T}}_{\xx,\xx'}(\omega) \hat{\psi}_k (\omega,\xx') d\xx' = \sigma_k^2(\omega) \hat{\psi}_k (\omega,\xx),
\end{equation}
where the eigenvalues $\sigma_k^2$ indicate the strength of the coherent (SPOD) mode $\hat{\psi}_k$. This tensor is computed with the Welch method where the expectation operator $\mathbb{E}[\cdot]$ is approximated by decomposing the signal of a single run (DNS or experimentally obtained) into $N_b$ overlapping and sufficiently long bins (of length $T$), leading to:
\begin{equation}
    \hat{\mathbf{T}}_{\xx,\xx'} (\omega) \approx \frac{1}{T} \frac{1}{N_b} \sum_{k=1}^{N_b} \mathcal{F}_T^k (\uu'(\xx')) (\omega) \mathcal{F}_T^{k,*} (\uu'(\xx)^*) (\omega) \equiv \hat{\mathbf{U}} \hat{\mathbf{U}}^{*,T},
\end{equation}
where $\mathcal{F}_T^k$ is the previously defined truncated Fourier transform, acting on the $k$-th bin, and $\hat{\mathbf{U}}(\omega) = [\cdots, \mathcal{F}_T^k (\uu') (\omega), \cdots ]/\sqrt{TN_b}$ is the (normalized) collection of Fourier-Transforms of all the bins. The superscript $^T$ denotes transposition. In a discrete in space system, the eigensystem becomes then:
\begin{equation}\label{eqn:SPOD2}
    \mathbf{U}(\omega) \mathbf{U}^*(\omega) B \hat{\psi}_k (\omega) = \sigma_k^2 (\omega) \hat{\psi}_k (\omega),
\end{equation}
where $B$ is a mass-matrix, containing the mesh-metrics, required by the space integral in \eqref{eqn:SPOD}. This eigensystem, if too large to be easily solved, can be transformed in a smaller one by considering the eigensystem for the variable $\hat{\mathbf{y}}_k$:
\begin{equation}\label{eqn:SPOD3}
    \hat{\mathbf{U}}^*(\omega) B \hat{\mathbf{U}}(\omega) \hat{\mathbf{y}}_k(\omega) = \sigma_k^2(\omega) \hat{\mathbf{y}}_k(\omega),
\end{equation}
and the SPOD mode can be recovered from $\hat{\psi}_k(\omega) = \sigma_k^{-1} \hat{\mathbf{U}}(\omega) \hat{\mathbf{y}}_k(\omega)$. Generally speaking, the dominant SPOD mode should be comparable with the dominant optimal response mode if the Resolvent operator is close to rank 1, that is $ \sigma_1  \gg \sigma_2 $ \citep{beneddine2016conditions}. If the expansion coefficients $\beta_k(\omega)$ are uncorrelated $\mathbb{E}[\beta_k \beta_l] = \alpha \delta_{k,l}$,
which is a strong requirement, then all SPOD modes should correspond to the optimal response modes \citep{towne2018spectral}.

\subsection{Numerical resolution of models}\label{sec:numsol}

In this paragraph, we describe the numerical method to solve the system of nonlinear equations given by \eqref{eqn:Model1} or \eqref{eqn:Model2}. One possible way is to iteratively solve the mean flow for a given dominant optimal response mode and then solve for a new mode with the new mean flow and keep alternating up to the point where a fixed point solution is reached. This method was employed by \cite{ManticLugo15} and \cite{ManticLugo16a} on a similar model. However, it was observed that for relatively high values of $|A|$, the method requires a low under-relaxation factor, making the convergence slow. Here we propose an alternative method, which consists in considering the set of equations \ref{eqn:Model1} as a whole and applying a Newton method to it. To do so, we first rewrite the eigensystem determining the dominant optimal response with the variables $(\lambda=\mu_1^2,\hat{\uu}_1,\hat{\mathbf{a}}_1=\lambda^{-1}\hat{\ff}_1)$:
\begin{equation}\label{eqn:SVD4}
	\begin{split}
		\lambda (\mathrm{i} \omega_0 B + L_{\tuu}) \hat{\uu}_1 & = B \hat{\mathbf{a}}_1, \;\;\; \hat{\uu}_1^{*,T} B \hat{\uu}_1 = 1 \\
         (\mathrm{i} \omega_0 B + L_{\tuu})^{*} \hat{\mathbf{a}}_1 & = B \hat{\uu}_1,
	\end{split}
\end{equation}
where $^*$ stands for complex-conjugation, $^T$ stands for vector-transposition and $B$ is the mass-matrix representing the energy scalar-product. We remark here that, for the sake of clarity, we keep the `continuous' notation introduced in previous paragraphs, even if now those objects are discrete, since we are dealing with the numerical method. Together with those equations, we need to solve the mean-flow equation as well (for simplicity, we only show the single-frequency case given by equation \ref{eqn:Model1}). Its linearization, together with the linearization of the rewritten eigensystem leads to:
\begin{equation}\label{eqn:NewtonSVD}
	\begin{split}
	\begin{bmatrix}
	L_{\tuu} & |A|^2 \phi ( \hat{\mathbf{u}}_1^* , (\cdot) ) + |A|^2 \phi ( \hat{\mathbf{u}}_1 , (\cdot)^* ) & 0 & 0 \\
	\lambda \phi ( (\cdot) , \hat{\mathbf{u}}_1 ) & \lambda (\mathrm{i} \omega_0 B + L_{\tilde{\mathbf{u}}}) & (\mathrm{i} \omega_0 B + L_{\tilde{\mathbf{u}}}) \hat{\mathbf{u}}_1 & - B \\ 
	0 & \hat{\mathbf{u}}_1^{*,T} B (\cdot) & 0 & 0 \\ 
	\phi^{*,T} ( (\cdot) , \hat{\mathbf{a}}_1 ) & - B & 0 & (\mathrm{i} \omega_0 B + L_{\tilde{\mathbf{u}}})^{*,T}
	\end{bmatrix}
    \begin{bmatrix}
	\delta \tilde{\mathbf{u}} \\ 
	\delta \hat{\mathbf{u}}_1 \\ 
	\delta \lambda \\ 
	\delta \hat{\mathbf{a}}_1
	\end{bmatrix} = \\
    \begin{bmatrix}
	- N(\mathbf{\tilde{u}},\mathbf{\tilde{u}}) - |A|^2 \phi (\hat{\mathbf{u}}_1, \hat{\mathbf{u}}_1^*) \\ 
	B \hat{\mathbf{a}}_1 - \lambda (\mathrm{i} \omega_0 B + L_{\tilde{\mathbf{u}}}) \hat{\mathbf{u}}_1 \\ 
	( 1 - \hat{\mathbf{u}}_1^{*,T} B \hat{\mathbf{u}}_1)/2 \\ 
	B \hat{\mathbf{u}}_1 - (\mathrm{i} \omega_0 B + L_{\tilde{\mathbf{u}}})^{*,T} \hat{\mathbf{a}}_1.
	\end{bmatrix}
	\end{split}
\end{equation}

This linear system is then solved with a GMRES iterative solver, preconditioned with the lower triangular matrix (lower Gauss-Seidel), completing the description of the numerical method used.

\section{An ``Oscillator" Flow - Squared-Section Cylinder} \label{sec:cylinder}

The two-dimensional flow around a squared-section cylinder is a typical example of oscillator flow. For Reynolds number larger than $Re \approx 50$ (see \cite{Sohankar98}), an unsteady behaviour naturally sets in, corresponding to a periodic limit cycle, characterized by a single frequency and its harmonics. Below, we will first (\S \ref{sec:cyl_conf}) present the details of the configuration and the numerical discretization. Then (\S \ref{sec:singlepoint}), we will present the results of the mean-flow and fluctuation reconstructions, tuned by a time-resolved measurement in the wake of the cylinder. 

\subsection{Configuration and numerical implementation}\label{sec:cyl_conf}

\begin{figure}
	\centering
	\includegraphics[trim={3cm 2cm 3cm 2cm},clip,width=200px]{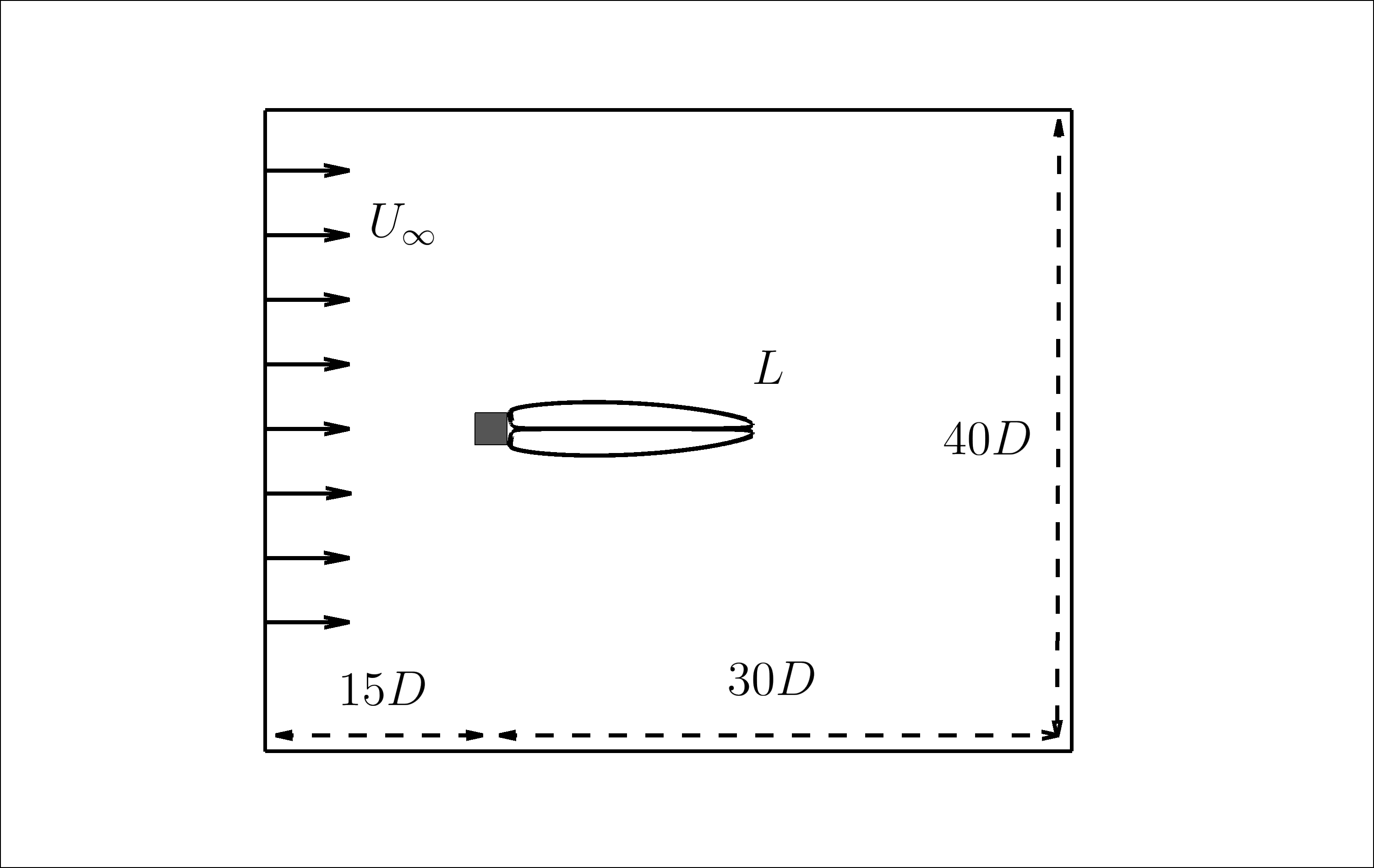}
    \caption{Sketch of the physical domain for the square cylinder case. The recirculation bubble's length is indicated in the figure and will be defined as $L$.}\label{fig:squaresketch}
\end{figure}

A sketch of the computational domain is provided in Figure \ref{fig:squaresketch}. The inflow boundary is located 15 diameters upstream of the center of the cylinder and a uniform velocity profile parallel to the lateral sides of the square is prescribed. The Reynolds number is set to $Re=D U_{\infty}/\nu = 100$. The lateral boundaries are located 20 diameters away from the center of the cylinder, on which we impose symmetry boundary conditions. The outflow boundary condition is imposed 30 diameters downstream of the cylinder and reads $(p I - Re^{-1} \nabla \mathbf{u}) \cdot \mathbf{n} = \mathbf{0}$. The reference flow, that will be probed, is obtained by time-averaging the solution of a Direct Numerical Simulation (DNS). The spatial discretization of both the DNS and the model relies on a Finite-Element Method (FEM) implemented in the FreeFEM++ code (see \citet{Hecht12}). We use second-order Taylor-Hood elements ($P^2$ for velocity and $P^1$ for the pressure fields). The DNS solver employs a second-order semi-implicit temporal scheme for time-advancement.

In Table \ref{tab:meshsquare}, we show various quantities of interest to characterize the mean- and unsteady features of the flowfield: the recirculation bubble's length $L$ (for the steady solution and the time-averaged unsteady solutions), the fundamental frequency $\omega_0$ (obtained via a Discrete-Fourier Transform on a sufficiently long DNS run) and the amplitude of the Fourier mode at the fundamental frequency in the unsteady simulation $|| \check{\uu}(\omega_0) ||$.
Results are provided for 3 meshes characterized by different grid densities. 
All of them are strongly refined around the square and in its wake and are coarsened in the free-stream region. We can see that mesh 2 provides a good agreement with the finer mesh 3 for all observed quantities. Mesh 2 will therefore be chosen as the default mesh in all following computations. In particular, the model will be computed on this mesh.
\begin{table}
	\centering
	\caption{Mesh dependency of steady and unsteady simulations for square cylinder at $Re=100$ - chosen mesh for the forthcoming computations in boldface. Observed parameters are recirculation bubble's length $L$, nonlinear frequency $\omega_0$, the norm of the first harmonic energy $||\check{\uu}(\omega_0)||$ and dominant gain, evaluated at the fundamental frequency of the flow $\mu_0^2 ( {\omega_0} )$.}\label{tab:meshsquare}
	\begin{tabular}{cccccccc} \hline
		 & & $\;\;$ & \multicolumn{2}{c}{Steady Solution} & \multicolumn{3}{c}{DNS} \\  
		 \cmidrule(lr){4-5}
		 \cmidrule(lr){6-8}
    	 & \# triangles & $\;\;$ & $L$ & $\mu_0^2 (\omega_0)$ & $L$ & $\omega_0$ & $||\check{\uu}(\omega_0)||$ \\ 
        \hline
   		Mesh 1 & 43000 & & 8.42 & 213775 & 2.43 & 0.913 & 2.131 \\
	    \textbf{Mesh 2} & 32000 & & 8.42 & 211745 & 2.43 & 0.912 & 2.130 \\
   	 	Mesh 3 & 12000 & & 8.40 & 206077 & 2.44 & 0.912 & 2.128 
	\end{tabular}
\end{table}

\subsection{Reference flow and data-assimilation results}\label{sec:singlepoint}

\begin{figure}
	\centering
    \raisebox{0.5in}{(a)}\includegraphics[trim={1.5cm 0cm 0cm 0cm},clip,height=70px]{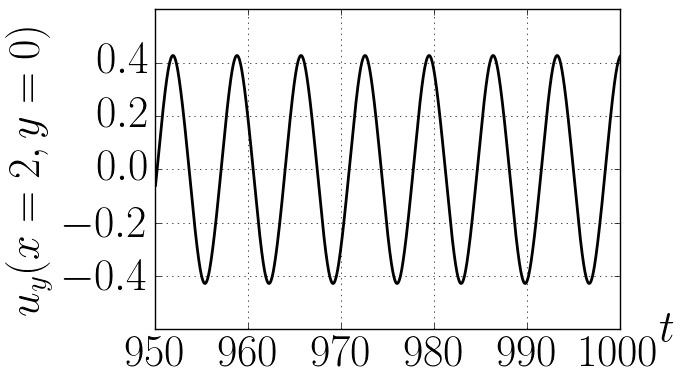}	
	\raisebox{0.5in}{(b)}\includegraphics[height=70px]{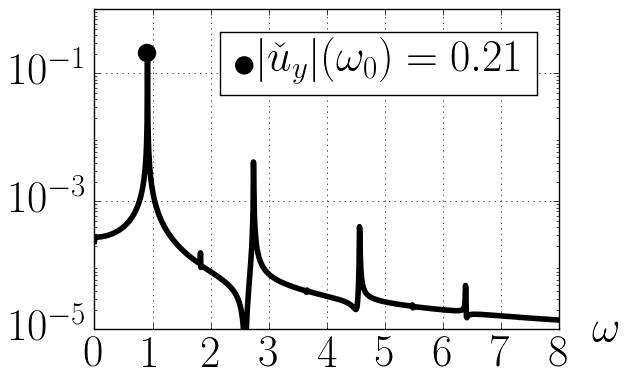}	
	\raisebox{0.5in}{(c)}\includegraphics[trim={1.5cm 0cm 0cm 0cm},clip,height=70px]{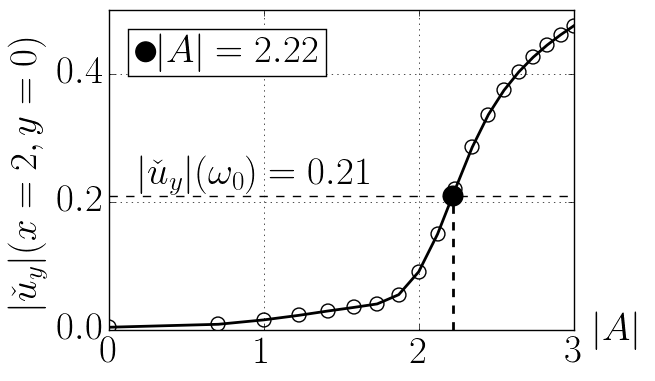}	
	\caption{Data-assimilation workflow for square cylinder flow: (a) measured signal corresponding to the cross-velocity $u_y$ at $(x,y) = (2,0)$, (b) Discrete-Fourier Transform of signal in (a), resulting in $\omega_{0} = 0.91$ and $|\check{u}_y|(\omega_0) = 0.21$. Tuning of $|A|$ in model \eqref{eqn:Model1} with this quantity (c), the optimal value being $|A| \approx 2.22$.} \label{fig:assimcylinder}
\end{figure}

Let us assume that we only know the time evolution of (for example) the cross-stream velocity component at a single point in the wake of the cylinder, for example  $\xx^m = (x,y) = (2,0)$.
This quantity is represented in Figure \ref{fig:assimcylinder}(a). The frequency content of this signal is computed with a Discrete-Fourier Transform, normalized such that it represents its harmonic-averages. In Figure $\ref{fig:assimcylinder}$ (b) we plot its modulus $|\check{u}_y|$ as function of frequency. This quantity indicates that the fundamental mode at $ \omega_0$ strongly dominates the harmonics $p\omega_0$ with $p\geq 2$. This suggests that a reconstruction with a single frequency may be sufficient. The fundamental frequency of the periodic limit cycles is equal to $\omega_0 = 0.91$, while the amplitude is $|\check{u}_y(\xx^m,\omega_0)| = 0.21$.

Since there is only one free parameter $|A|$ in model \eqref{eqn:Model1},
we may explore the control parameter space by computing the reconstructed solution for a range of $|A|$ and therefore tuning $|A|$ with the reference measurement data shown in figures \ref{fig:assimcylinder}(a,b). From a numerical point of view, this does not necessarily represent a high computational effort since one may proceed by progressively increasing the value of $|A|$ and restart each new computation from the previous solution (for $|A|=0$, the reconstructed mean-flow solution is the base-flow). Changing the value of $|A|$ in sufficiently small steps, the reconstructed solution is obtained in a few Newton iterations. As $|A|$ increases, the number of GMRES iterations to invert eq. \eqref{eqn:NewtonSVD} increases since the preconditioner becomes less accurate. 

\begin{figure}
	\centering
    \raisebox{0.3in}{(a)}\includegraphics[trim={1cm 1cm 1cm 14cm},clip,width=160px]{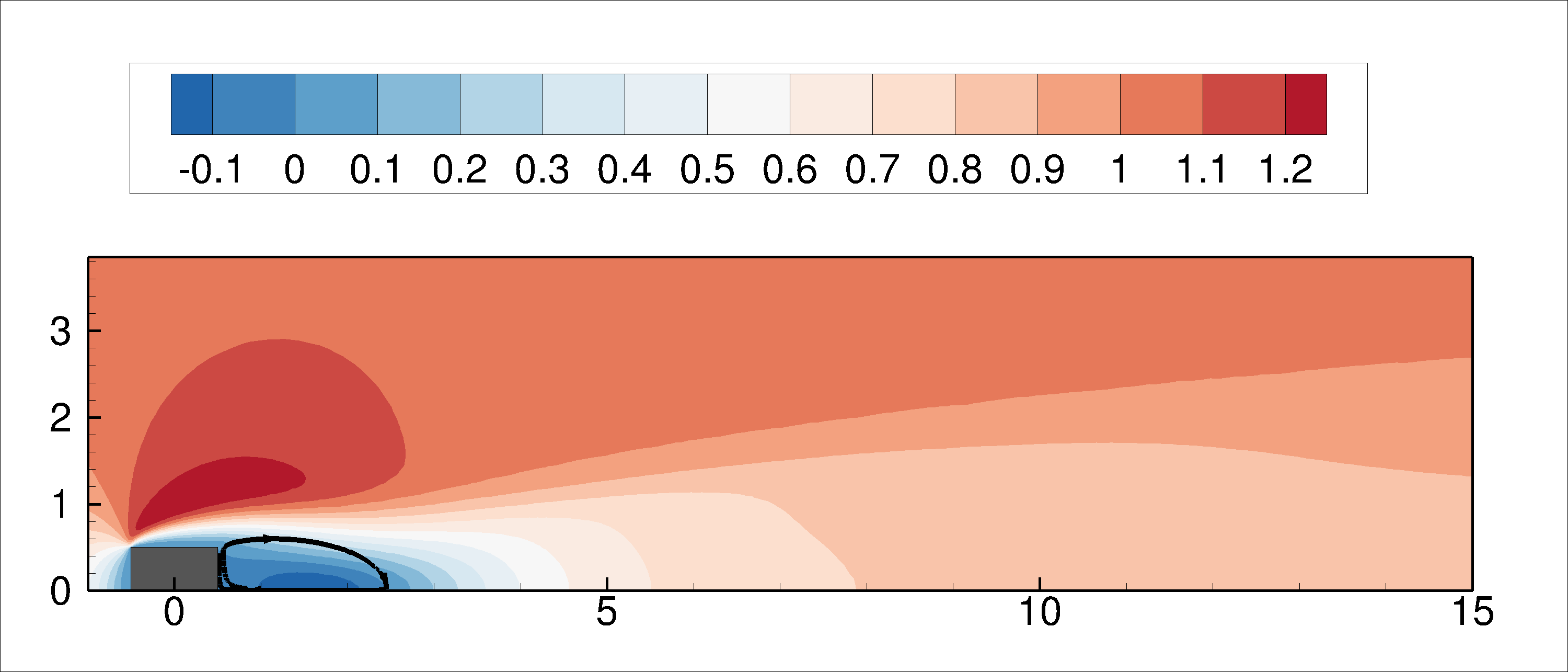}
	\raisebox{0.3in}{(b)}\includegraphics[trim={1cm 1cm 1cm 14cm},clip,width=160px]{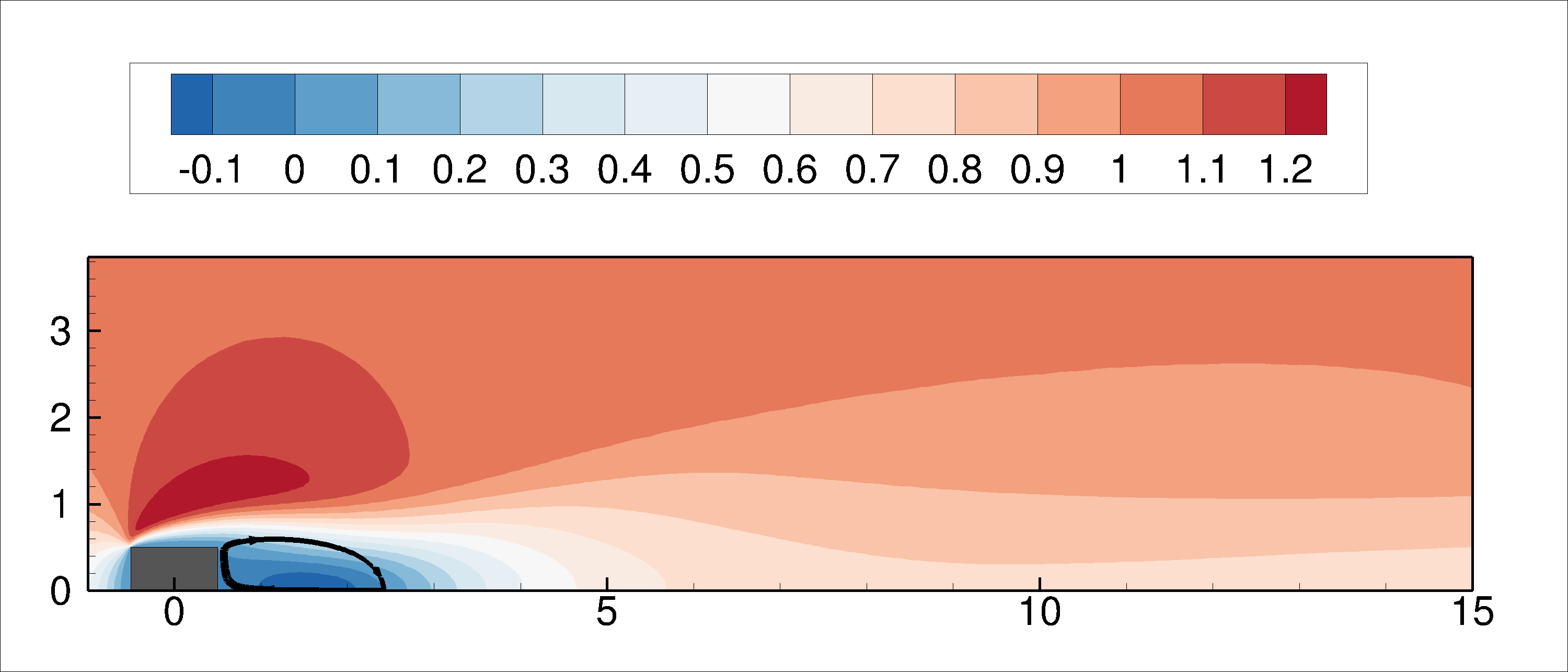}
  
    \raisebox{0.3in}{(c)}\includegraphics[trim={1cm 1cm 1cm 14cm},clip,width=160px]{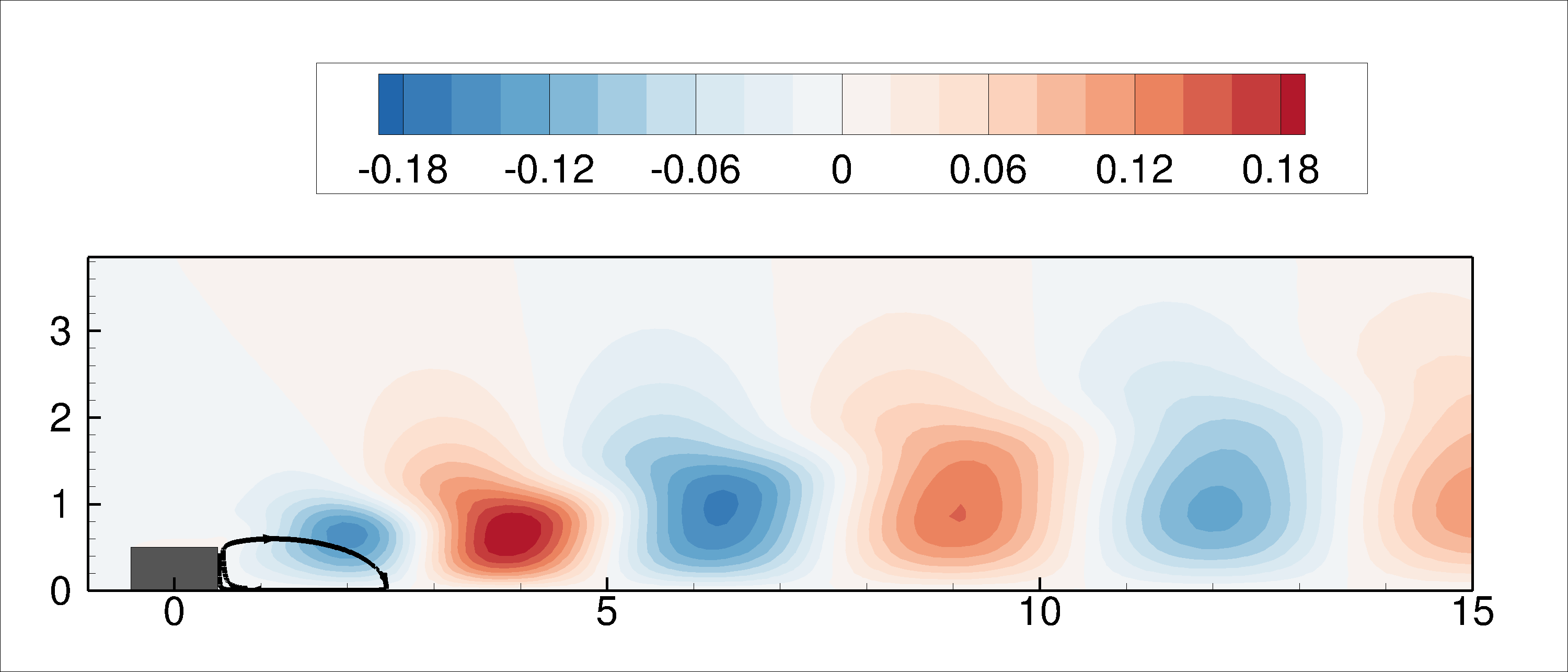}
	\raisebox{0.3in}{(d)}\includegraphics[trim={1cm 1cm 1cm 14cm},clip,width=160px]{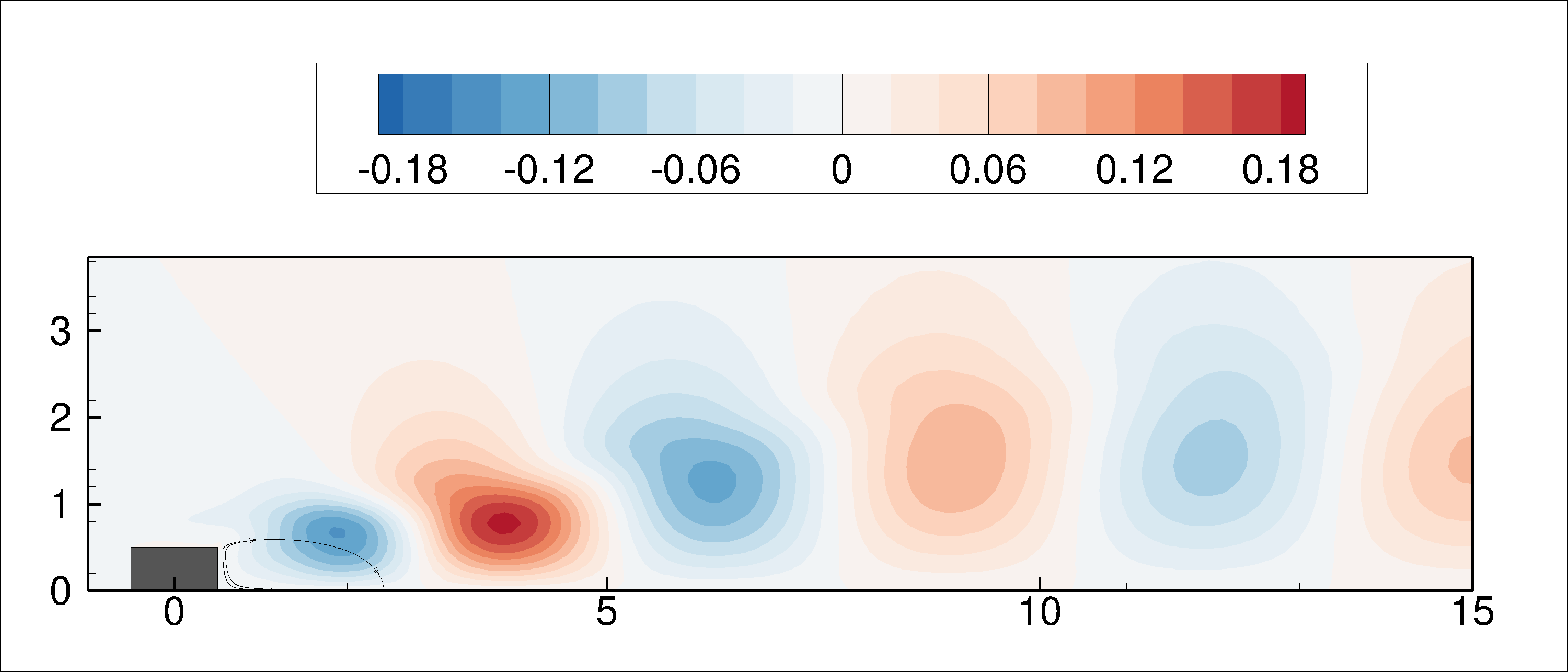}
    
    \raisebox{0.3in}{(e)}\includegraphics[trim={1cm 1cm 1cm 14cm},clip,width=160px]{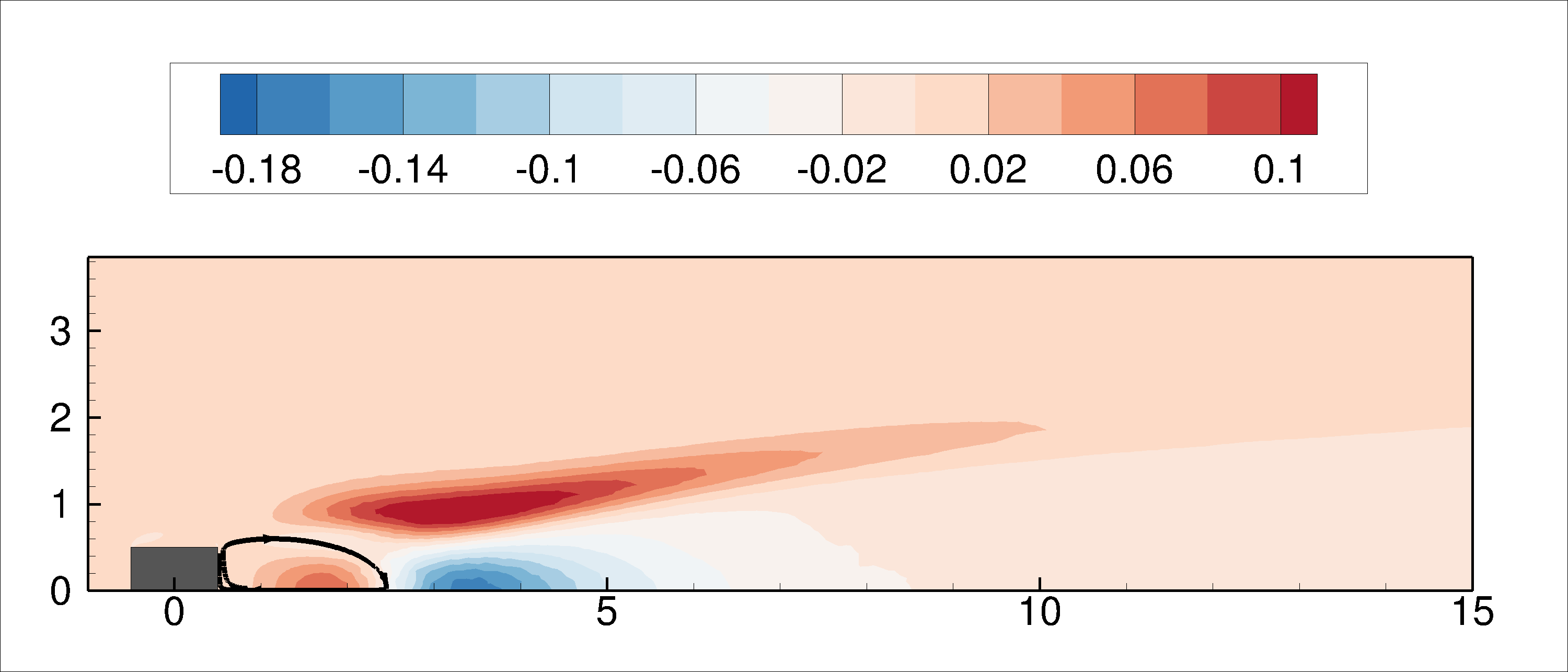}
	\raisebox{0.3in}{(f)}\includegraphics[trim={1cm 1cm 1cm 14cm},clip,width=160px]{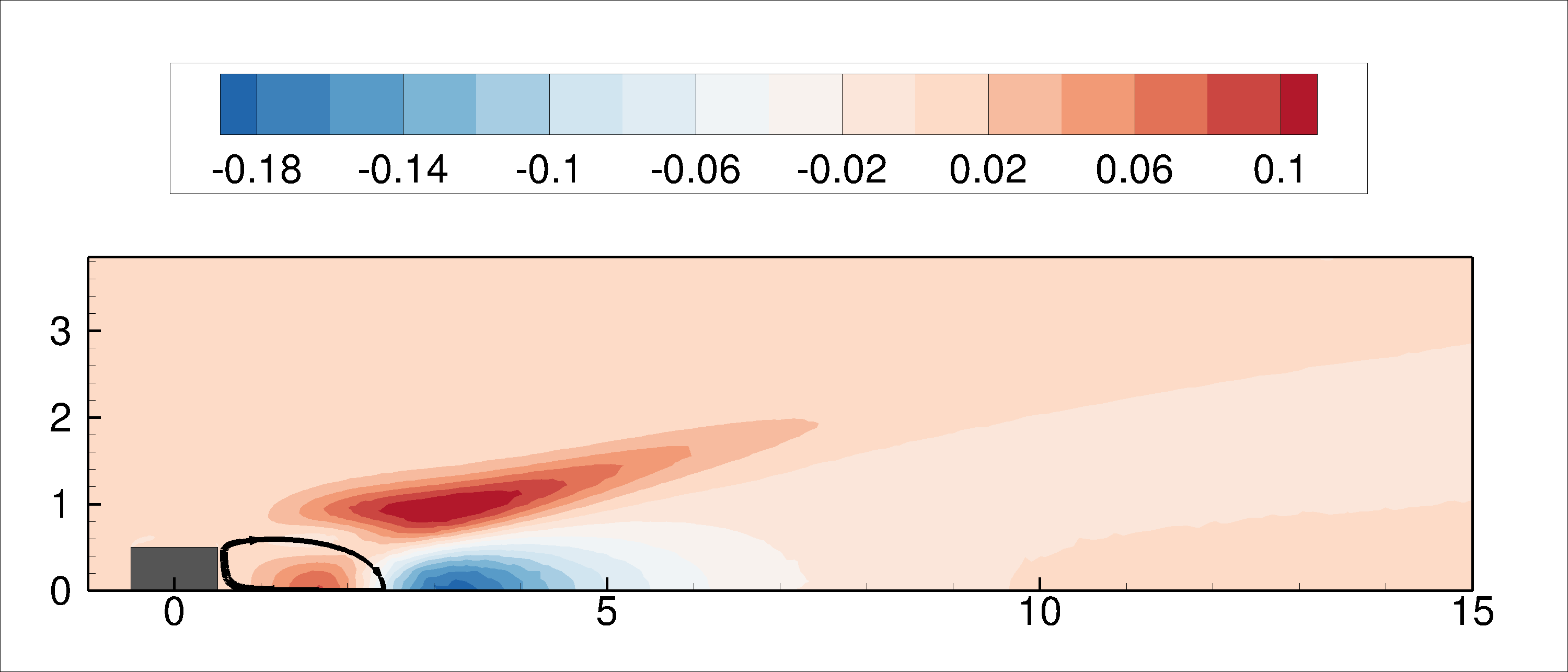}

	\caption{Comparison between reconstructed flow features (left column) and DNS (right column) for the square-cylinder configuration at $ Re=100$. Stream-wise component of (a,b) mean-flow velocities (c,d) first-harmonic streamwise velocities $A \hat{\mathbf{u}}_0$ and $\check{\uu}(\omega_0)$ and (e,f) mean force induced by the Reynolds stress tensor $\tilde{\mathbf{f}} = 2 |A|^2 \bnabla \cdot \Re \{ \hat{\mathbf{u}}_1 \otimes \hat{\mathbf{u}}_1^* \}$  with $|A|=2.22$. The black curves delimit the recirculation regions of the reconstructed mean flow.} \label{fig:reconstrucutedunsteadyflows100}
\end{figure}

\begin{figure}
	\centering
	\raisebox{0.7in}{(a)}\includegraphics[width=105px]{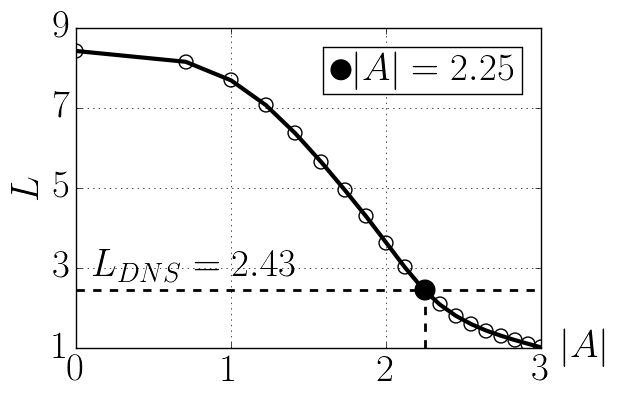}
	\raisebox{0.7in}{(b)}\includegraphics[width=110px]{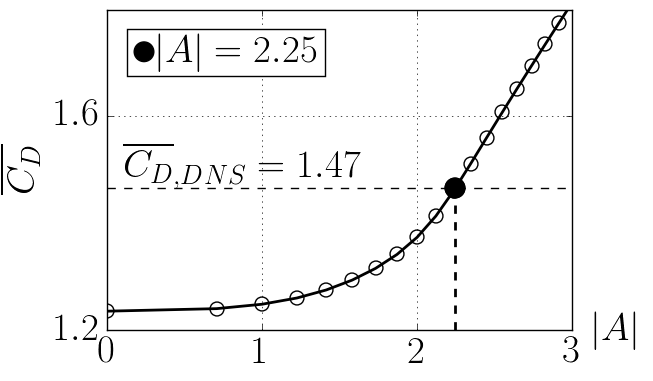}
	\raisebox{0.7in}{(c)}\includegraphics[width=105px]{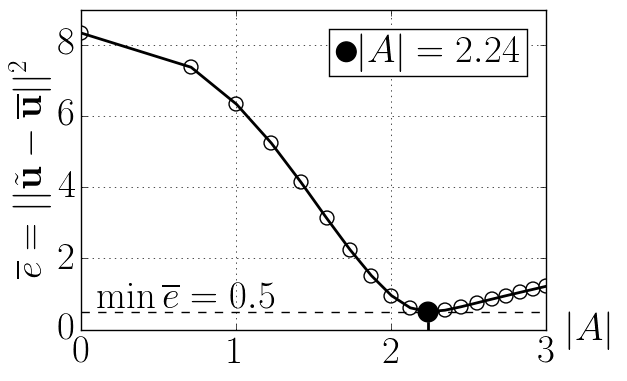}
    
    \caption{Reconstruction of other quantities: (a) the length of the recirculation bubble, (b) Mean drag coefficient (c) global mean-flow error $\overline{e} = \left \| \tilde{\mathbf{u}} - \overline{\mathbf{u}} \right \|^2$. For each different measure, we may have a different (but close) value of the estimated $|A|$.} \label{fig:quantities-A}
\end{figure}

Once the solutions of the model are obtained, we compare the prediction $|A| |\hat{\uu}(\xx^m,\omega_0)|$ (shown with a solid line in figure \ref{fig:assimcylinder}(c)) with the measurement $|\check{u}_y|(\xx^m,\omega_0)$ (horizontal dashed-line). We remark that the predicted quantity is always increasing with $|A|$, meaning that there exists a unique value of $|A|$ for which the predicted and measured values match. This procedure leads to a value of $|A|=2.22$. This value slightly overestimates the actual DNS values, since $||\check{\uu}(\omega_0)||=2.13$. This is due to the fact that the model only takes into account the fundamental harmonic and not the higher order harmonics. The effect of the latter modes on the mean-flow distortion is compensated by a slight overestimation of the amplitude of the fundamental Fourier mode. However, despite this small overestimation of the mode's energy, the reconstructed solution of the model \ref{eqn:Model1}, represented in Figure \ref{fig:reconstrucutedunsteadyflows100} by the mean-flow (a), Fourier mode (b) and Reynolds-stress divergence (c), compares very well with the reference counterparts (b,d,f). Furthermore, if we consider other flow quantities (such as the recirculation length $L$, the mean-drag $\overline{C_D}$ or the error related to the mean-flow estimation $\overline{e}=||\tilde{\uu}-\ouu||^2$) to tune $ |A|$, we can see in Figure \ref{fig:quantities-A}
that all of them yield a similar value $|A|$, between $2.22$ and $2.25$. This shows the robustness of the approach for this particular flow, which stands as an appealing alternative to the more classical mean-flow data-assimilation approaches \citep{Foures14}, especially if only few scalar measurements are available.


At this point, we remind that a similar model was proposed by \cite{ManticLugo15} for the flow around a circular cylinder. Two major differences can be highlighted.  First, instead of using optimal response modes, they use global modes for the approximation of the harmonics. Second, instead of considering external data to provide a criterion for the choice of $|A|$, their model searched for a value of $|A|$ such that the eigenmode was marginally stable. This condition is justified for flows behaving harmonically, where higher order harmonics are negligible with respect to the fundamental. Hence, their method is almost exclusive to harmonic "oscillator" flows, where unsteadiness is triggered by intrinsic dynamics. In the next section, we show the generality of our approach by considering the case of an "amplifier flow", here a backward-facing step flow where the fluctuation is triggered by external noise, amplified through linear convective instabilities (\cite{Marquet10}).

\section{A ``Noise-Amplifier" Flow - Backward-facing step}\label{sec:BFS}

In this section, we consider the (more challenging) backward facing step flow configuration, which exhibits a broadband behaviour. We first briefly present the configuration (\S \ref{sec:BFS_conf}) and then assess the reconstruction procedure.

\subsection{Configuration}
\label{sec:BFS_conf}

\begin{figure}
	\centering
	\includegraphics[trim={0.5cm 0.5cm 0.5cm 0.5cm},clip,width=400px]{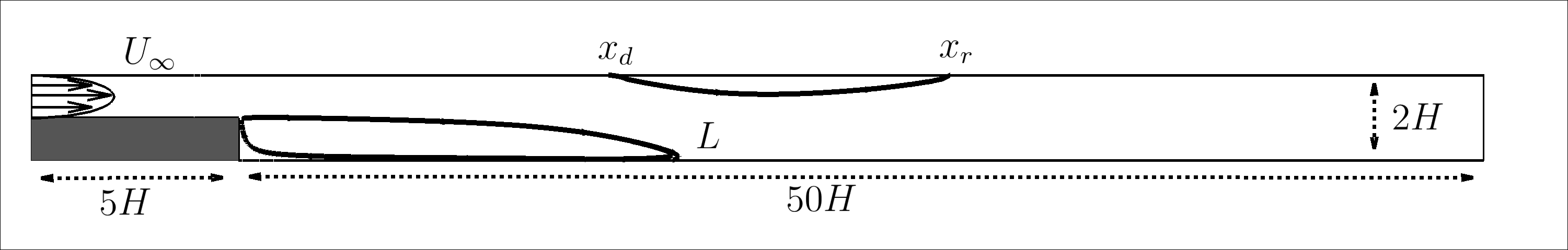}
    \caption{Sketch of the physical domain and inflow boundary condition for the backward-facing step case. The length of the main recirculation bubble, the detachment and reattachment points at the top wall are indicated in the figure and defined to be $L$, $x_d$ and $x_r$, respectively.} \label{fig:stepsketch}
\end{figure}

The configuration is the Backward-facing step described in \cite{Herve12,Barkley02}. The Reynolds number, based on the height of the step $H$ and the maximum inlet velocity $ U_\infty$ of the Poiseuille flow imposed at the inlet, is fixed at $Re=U_{\infty} H/\nu=500$. The inflow boundary condition is located at $5H$ upstream of the step and the outflow is located at $50H$ downstream, where, again, the boundary condition $(p I - Re^{-1} \nabla \mathbf{u}) \cdot \mathbf{n} = \mathbf{0}$ is imposed. The other boundaries of the domain correspond all to solid walls on which a no-slip boundary condition is imposed. A sketch of the computational domain is given in Figure \ref{fig:stepsketch}.

For the present flow conditions, there exists a globally-stable steady-state solution (see \cite{Barkley02}, \cite{Blackburn08}) which exhibits strong linear selective-frequency amplification mechanisms (\citet{Marquet10}). It can be triggered by considering upstream stochastic noise, whose amplitude is tuned so that the dynamics of the perturbations is nonlinear and generates a mean-flow deformation. In the present study, we pick that forcing term to be:
\begin{equation}
    \mathbf{f}(\xx,t) = C \left( \begin{array}{c} 0 \\ \phi(\xx) \end{array} \right) w(t),
\end{equation}
where $C=10$ is the amplitude of the signal (chosen sufficiently high so that nonlinear effects are effective), $w(t)$ is a (pseudo) white-noise and $\phi(\mathbf{x}) = e^{-|\mathbf{x}-\mathbf{x}_G|^2/2\sigma^2}$ is a Gaussian function centered at $\mathbf{x}_G = (-0.5,0.25)$ of width $\sigma = 0.1$ (for further details, see \citet{Herve12}). In table \ref{tab:meshstep}, we provide some brief mesh-convergence results in terms of steady-solution and unsteady time-averaged quantities, which are resolved through a two-dimensional second-order in time DNS (see \citet{sipp2016linear,Herve12}, with $\Delta t=0.005$). We can see that for both meshes the observed quantities are similar, with a difference of up to $\sim 5\%$. For this reason, we found reasonable to keep the coarser mesh (Mesh 2) as the default mesh for all subsequent computations as a good compromise of accuracy and computational time.

\begin{table}
	\centering
    \caption{Mesh dependency of steady and time-averaged unsteady simulations for backward-facing step in terms of recirculation quantities ($L$, $x_d$, $x_r$), and fluctuation energy for, the unsteady solution $\mathcal{A}=\sqrt{\frac{1}{T} \int_0^T |\uu'|^2}$. Chosen mesh for the computations in boldface}
	\begin{tabular}{cccccccccc} \hline
		 & & \multicolumn{3}{c}{Steady Solution} & \multicolumn{5}{c}{Unsteady Solution} \\ 
        \cmidrule(lr){3-5}
        \cmidrule(lr){6-9}
    	 & \# triangles & $L$ & $x_d$ & $x_r$ & $L$ & $x_d$ & $x_r$ & $\mathcal{A}$ \\
    	\hline
   		Mesh 1 & 35000 & 10.875 & 8.688 & 17.508 & 7.120 & 5.781 & 9.912 & 1.07 & \\
    	\textbf{Mesh 2} & 17000 & 10.870 & 8.685 & 17.508 & 7.461 & 5.761 & 10.405 & 1.07 & 
	\end{tabular} \label{tab:meshstep}
\end{table}

\subsection{Reference flow and data-assimilation results}
\label{sec:broadband}

\begin{figure}
	\centering
	\raisebox{0.3in}{(a)}\includegraphics[trim={0cm 0cm 0cm 0cm},clip,height=55px]{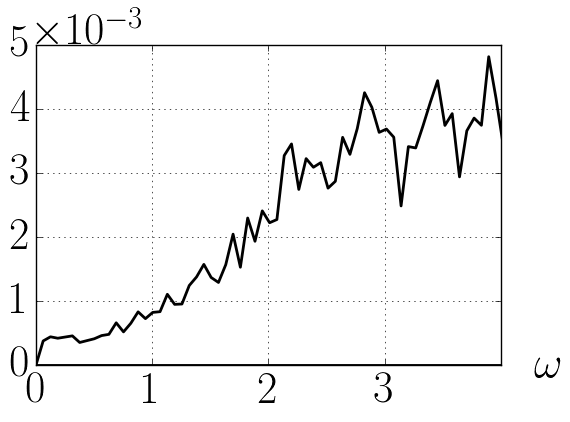} 
	\raisebox{0.3in}{(b)}\includegraphics[trim={0cm 0cm 0cm 0cm},clip,height=55px]{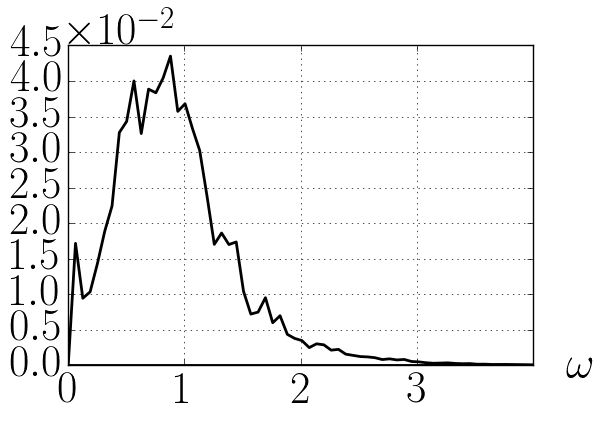}
	\raisebox{0.3in}{(c)}\includegraphics[trim={0cm 0cm 0cm 0cm},clip,height=55px]{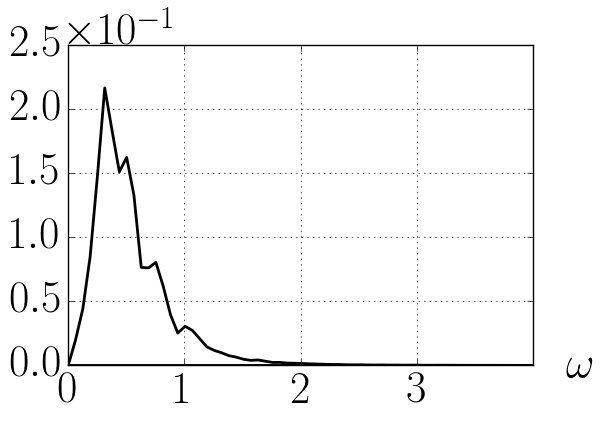}
	\raisebox{0.3in}{(d)}\includegraphics[trim={0cm 0cm 0cm 0cm},clip,height=55px]{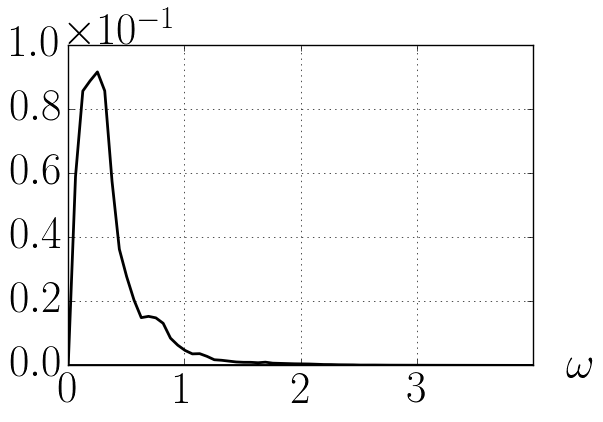}
    \caption{DNS results for white-noise external forcing: Power-Spectral Density (PSD) of the cross-stream velocity at (a) $\xx_p^1 = (0,0.2)$, (b) $\xx_p^2 = (4,0.2)$, (c) $\xx_p^3 = (10,0.2)$ and (d) $\xx_p^4 = (15,0.2)$.}
	\label{fig:unsteadyBFS-wn}
\end{figure}

Similarly to what was done in the time-periodic case, we start by considering time-series of signals at a few locations of the domain. This shows roughly how the energy of the fluctuations is distributed in the domain, especially as a function of frequency. In Figure \ref{fig:unsteadyBFS-wn}, we show the PSD (computed using the Welch method, considering a time-series of length 5000 time units, subdivided in 99 bins with 50\% overlap, with a sample time of $\Delta t = 0.005$) for locations spanning from the vicinity of the step location (a), to the ouflow boundary (b,c,d). We can see from those signals that only at the very vicinity of the step (and thus close to the location where the external white-noise forcing is applied) the signal contains high frequencies. For the signals extracted further downstream, the frequency range is restricted mainly to $\omega \in (0,2)$ (figures b,c,d). Furthermore, the amplitude of those spectra is much higher than in (a), indicating that the dynamics at these frequencies stems from convective amplification mechanisms linked to the shear-layer generated at the step. This is in accordance with \citet{Marquet10}, who showed, thanks to a Resolvent analysis, that the peak of energy amplification is around $\omega \approx 0.5$. For this reason, for our model, we will choose $\Omega_{N+1}=2$ and a uniform frequency discretization of $\Delta \omega = \Omega_{j+1}-\Omega_j$, for which $\omega_j=\Delta \omega (j+1/2)$. 

As done previously, we may compare the model-predicted energy of the fluctuation at points $\xx_p^m$ to the ones measured. The associated error measure is given by:
\begin{equation}
    J_{\omega_j,\xx_p^m} = \left( \frac{2 \pi |A_j|^2 |\hat{\mathbf{u}}_0^j (\xx_p^m)|^2}{I_{\xx_p^m,\omega_j}} - 1 \right)^2.
\end{equation}
We are now able to define our global cost functional, as a sum over at least $N$ of those ``elementary" contributions $J_{\omega_j,\xx_p^m}$, possibly choosing different points $\xx_p^m$ and different frequencies $\omega_j$. In what follows, we will choose always $N$ of those quantities so as to be able to tune the $N$ control parameters $ |A|_j $. For robustness reasons, we pick points $\xx_p^m$ and frequencies $ j $ so that fluctuation energy is high in the frequency band linked to $\omega_j$. For this reason, we will ignore the signal at $\xx_p^1$ where the fluctuation level is weak for all frequencies. Furthermore, we can see that the signal $\xx_p^2$, contrarily to $\xx_p^{3,4}$ contains a significant amount of energy within $1\leq\omega<2=\Omega_{N+1}$. For this reason, we will ``trust" this signal in that frequency range. For the range $\Omega_0=0\leq\omega<1$, we will use either of the two signals $\xx_p^{3,4}$.
These choices result in the following 2 cost-functionals that will be minimized:
\begin{equation}
    J^1 = \sum_{\omega_j\geq1} J_{\omega_j,\xx_p^{2}}+\sum_{\omega_j<1} J_{\omega_j,\xx_p^{3}}, \;\;\;\; J^2 =
    \sum_{\omega_j\geq1} J_{\omega_j,\xx_p^{2}}+\sum_{\omega_j<1} J_{\omega_j,\xx_p^{4}} .
\end{equation}

We will consider different numbers $N$ of frequency bands and analyse the effect of this parameter on the reconstructed field. We will consider the cases $N=2,5,10$. Also, to speed up the convergence for the cases $N=5$ and $10$, we first perform the optimization for the case $N=2$, which is quick.
Then, we initialize the mean-flow for $ N=5$ with the solution obtained for $N=2$ and the $|A_j|$ coefficients by computing the Resolvent modes (again based on the mean-flow determined with $N=2$) and setting $|A_j| = \sqrt{I_{\mathbf{x}_p^m,\omega_j}/2 \pi} / | \hat{\mathbf{u}}^j_1 (\mathbf{x}_p^m) | $, where $\xx_p^m$ is the current measurement point. We repeat the same procedure to initialize the optimization for $N=10$ based on the mean-flow obtained for $ N=5$.

\subsubsection{Reconstruction results}

\begin{figure}
	\centering
    \raisebox{0.5in}{(a)}\includegraphics[height=100px]{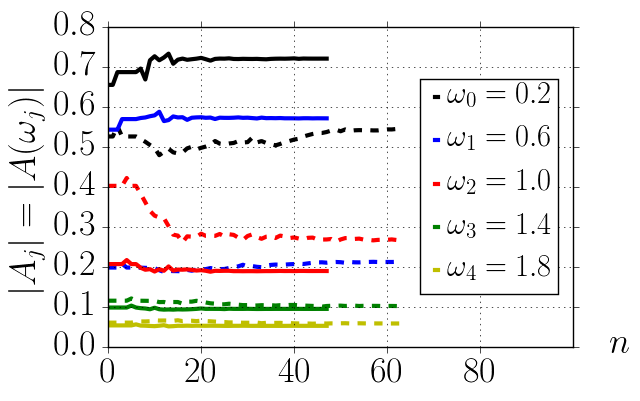}    
	\raisebox{0.5in}{(b)}\includegraphics[height=100px]{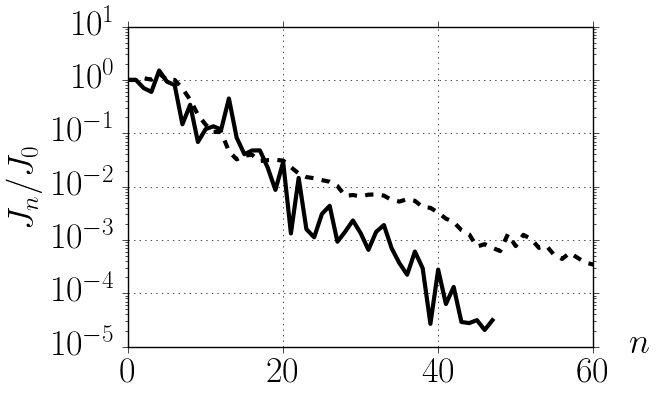}

    \caption{Convergence of the amplitudes $|A_j|$ and of the cost functional $J_n/J_0$ for $J=J^1$ (solid lines) and $J=J^2$ (dashed lines). Case $N=5$}\label{fig:COBYLA}
\end{figure}

In figure \ref{fig:COBYLA} we represent the convergence of the optimization process for the case $N=5$ and for the two cost functionals $ J^1 $ and $ J^2 $ described above. We can see that starting from the case $N=2$ provides indeed a good initial guess, especially for $J^1$, since  the optimization did not drift apart too much from the initialization and the optimisation did not take more than 20 iterations to converge up to a precision of $J_n/J_0 \approx 10^{-2}$. For the cost functional $J^2$, the optimization needed slightly more iterations, mainly because of a poor estimate of the initial amplitude for $\omega_2=1$. Furthermore, we observe that the estimate of the amplitude for $\omega_0=0.2$ and $\omega_1=0.6$ is considerably lower than those predicted for  $J^1$. The rate of convergence of the cost functional $J_n/J_0$ is almost exponential for both cost functionals.

\begin{table}
	\centering
    \caption{Reconstructed parameters in function of number of frequencies considered $N$.}
	\begin{tabular}{ccccccccccc}
		 & \multicolumn{2}{c}{$L$} & \multicolumn{2}{c}{$x_d$} & \multicolumn{2}{c}{$x_r$} & \multicolumn{2}{c}{$\overline{e} = || \tuu - \ouu ||^2$} & \multicolumn{2}{c}{$\mathcal{A} = \sqrt{ \frac{1}{T} \int_0^T \int_{\Omega} |\uu'|^2 }$} \\
		\cmidrule(lr){2-3}
        \cmidrule(lr){4-5}
        \cmidrule(lr){6-7}
        \cmidrule(lr){8-9}
        \cmidrule(lr){10-11}
	   	DNS & \multicolumn{2}{c}{$7.07$} &  \multicolumn{2}{c}{$5.62$} &  \multicolumn{2}{c}{$9.99$} &  \multicolumn{2}{c}{--} &  \multicolumn{2}{c}{$1.07$} \\  
        \cmidrule(lr){2-3}
        \cmidrule(lr){4-5}
        \cmidrule(lr){6-7}
        \cmidrule(lr){8-9}
		\cmidrule(lr){10-11}
		 & $J^1$ & $J^2$ & $J^1$ & $J^2$ & $J^1$ & $J^2$ & $J^1$ & $J^2$ & $J^1$ & $J^2$ \\ 
        \cmidrule(lr){2-3}
        \cmidrule(lr){4-5}
        \cmidrule(lr){6-7}
        \cmidrule(lr){8-9}
        \cmidrule(lr){10-11}
		$N=2$  & 7.42 & 9.39 & 5.91 & 7.57 & 9.31 & 13.34 & 0.10 & 0.70 & 1.20 & 0.81 \\  
		$N=5$  & 6.25 & 8.67 & 5.24 & 6.93 & 8.16 & 12.01 & 0.05 & 0.40 & 1.33 & 0.92 \\  
		$N=10$ & 6.02 & 7.78 & 5.35 & 6.77 & 7.80 & 10.40 & 0.09 & 0.20 & 1.41 & 1.11 
	\end{tabular} \label{tab:whitenoise}
\end{table}

\begin{figure}
	\centering
    \raisebox{0.5in}{(a)}\includegraphics[trim={1.15cm 0cm 1cm 0cm},clip,height=80px]{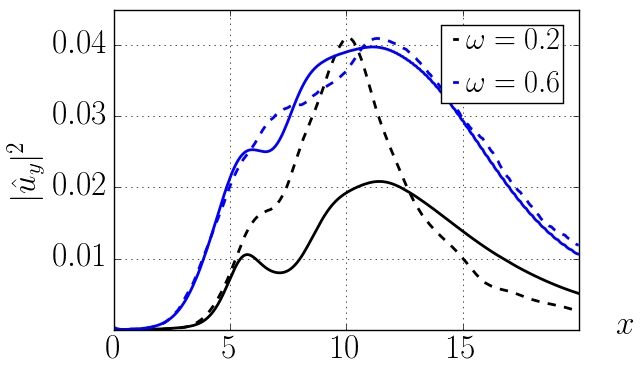}    
	\raisebox{0.5in}{(b)}\includegraphics[trim={1.15cm 0cm 1cm 0cm},clip,height=80px]{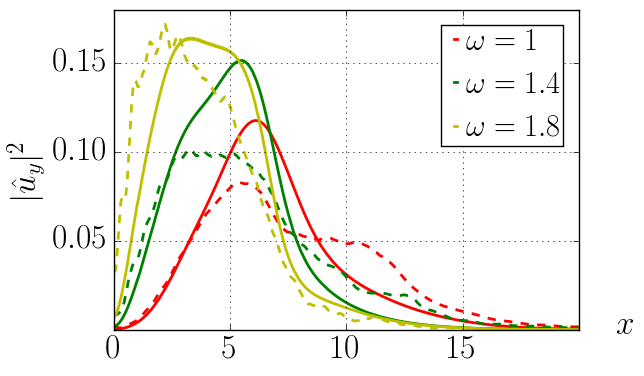}
    \caption{Spatial distribution of the energy, evaluated on the line $y=0.2$, of the resolvent modes on the mean flow (solid lines) and Fourier modes (dashed lines) at several frequencies (both modes are normalized).}\label{fig:energy-resolvent-mean-white-noise}
\end{figure}

In table \ref{tab:whitenoise}, we have compared, for the 2 cost functionals and 3 numbers of frequencies ($ N =2,5,10 $), some overall features of the reconstructed flow with those of the DNS. We first remark that, if we take an increasing number of modes in the model ($N$), we resolve better the frequency interval and we tend to add more energy into the system. This energy can be measured by the quantity $\mathcal{A} = \sqrt{ \frac{1}{T} \int_0^T \int_{\Omega} |\uu'|^2 }$, homogeneous to an amplitude, which for the model we use can be recast as $\mathcal{A} = \sqrt{\sum_j 2 |A_j|^2}$. We can also see that, for the cost functional $J^2$, this value gets closer to the exact one (computed from the DNS snapshots). This is not the case for the first cost functional, which over-predicts this quantity, especially as $N$ increases. We can thus see that, although the measurement points were chosen so that their signals contain a significant and relevant amount of energy, they predict slightly different scenarios. A possible explanation of this fact can be given in figure \ref{fig:energy-resolvent-mean-white-noise} where we plot the energy of the resolvent modes and of the SPOD modes along the measurement line $y=0.2$ as a function of the stream-wise coordinate $x$, showing that, for example, at $\omega = 0.2$, the SPOD mode (dashed line) is stronger than the resolvent mode (solid line) for $x<12$ and the other way round for $ x>12.5$, suggesting that if the signal from $\xx_p^3$ ($x=10$) is used (which is the case for $J^1$), the energy for this frequency is probably going to be overpredicted, increasing $\mathcal{A}$. On the other hand, if $\xx_p^4$ ($x=15$) is used ($J^2$) this energy will tend to be more accurately predicted.

Looking now at the mean-flow quantities, such as the detachment / reattachment lengths $L,x_d,x_r$, we can see that, by increasing $N$, and consequently $\mathcal{A}$, we have almost always a shortening of those lengths, with an exception for $x_d$ with $J^1$, where it stayed at a constant value, near the reference one. This shows that, generally, the more fluctuation energy is provided to the system, the shorter the bubbles. However, those quantities alone are not capable to infer, in a conclusive manner, how well the mean-flow is reconstructed. For this reason, we evaluate as well the mean-flow error $\overline{e}=|| \tilde{\uu} - \overline{\uu} ||^2$. We can see that, although the second cost functional $J^2$ provides a better prediction of the fluctuation energy $\mathcal{A}$, its mean-flow prediction is poorer than with $J^1$, for which the mean-flow error is of the order of $\overline{e} \in (0.05,0.1)$ (the minimal value being observed for $N=5$), against $\overline{e} \in (0.2,0.7)$ for $J^2$. This interplay between a good prediction of the energy of the fluctuation field and a good mean-flow prediction was already observed in the previous cylinder configuration (although much more mitigated) where the value of $A$ minimizing the mean-flow error $\overline{e}$ was higher than the DNS one due to the absence in the model of the second-order harmonic. A similar argument can therefore be made here.

\begin{figure}
	\centering
    \raisebox{0.2in}{(a)}\includegraphics[trim={1cm 1cm 1cm 1cm},clip,width=300px]{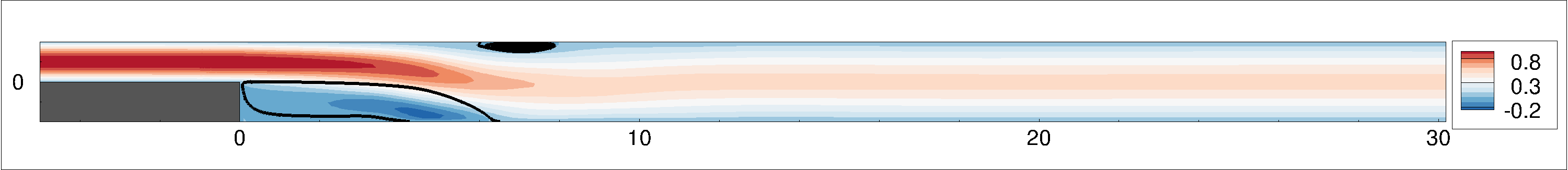}
    \raisebox{0.2in}{(b)}\includegraphics[trim={1cm 1cm 1cm 1cm},clip,width=300px]{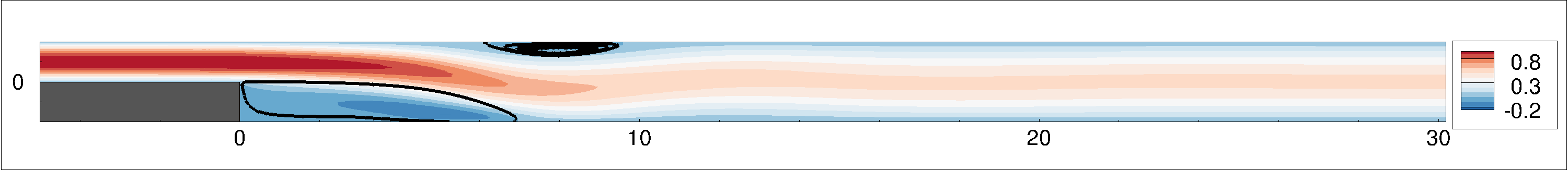}
    
    \raisebox{0.2in}{(c)}\includegraphics[trim={1cm 1cm 1cm 1cm},clip,width=300px]{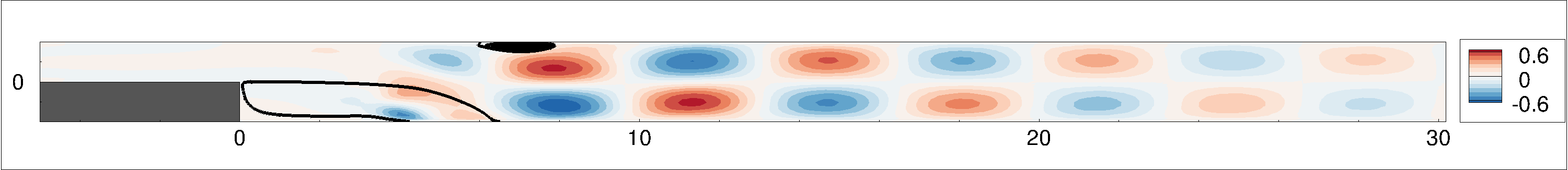}
    \raisebox{0.2in}{(d)}\includegraphics[trim={1cm 1cm 1cm 1cm},clip,width=300px]{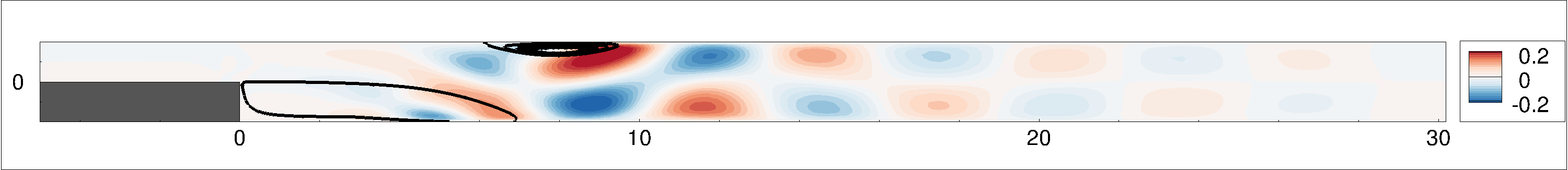}

    \raisebox{0.2in}{(e)}\includegraphics[trim={1cm 1cm 1cm 1cm},clip,width=300px]{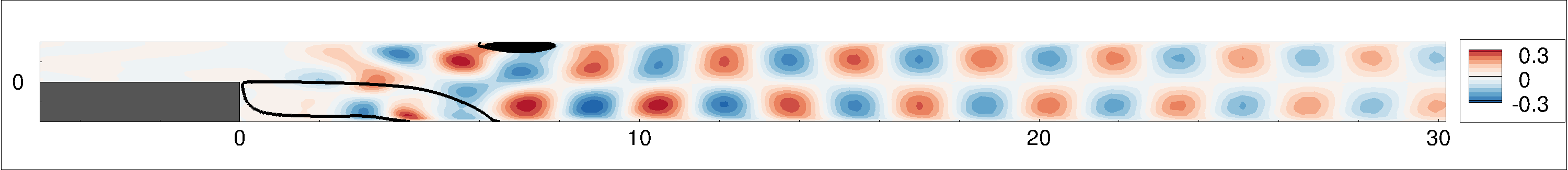}
    \raisebox{0.2in}{(f)}\includegraphics[trim={1cm 1cm 1cm 1cm},clip,width=300px]{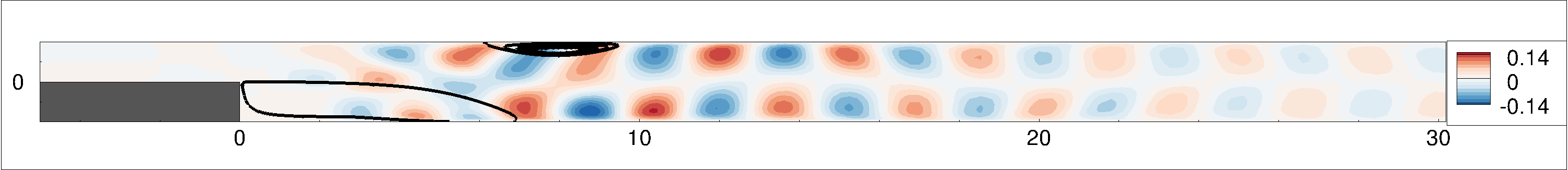}
        
    \raisebox{0.2in}{(g)}\includegraphics[trim={1cm 1cm 1cm 1cm},clip,width=300px]{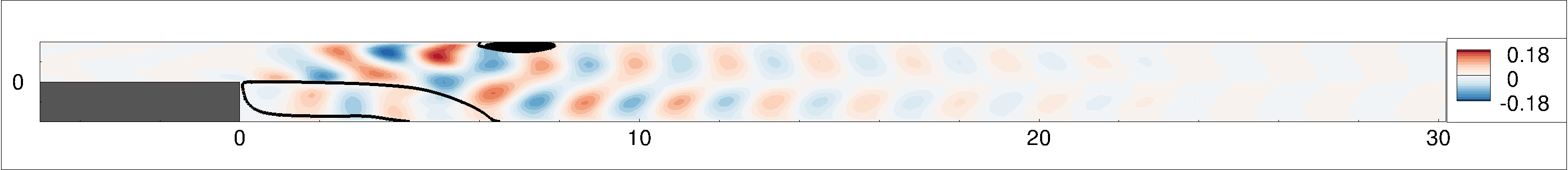}
    \raisebox{0.2in}{(h)}\includegraphics[trim={1cm 1cm 1cm 1cm},clip,width=300px]{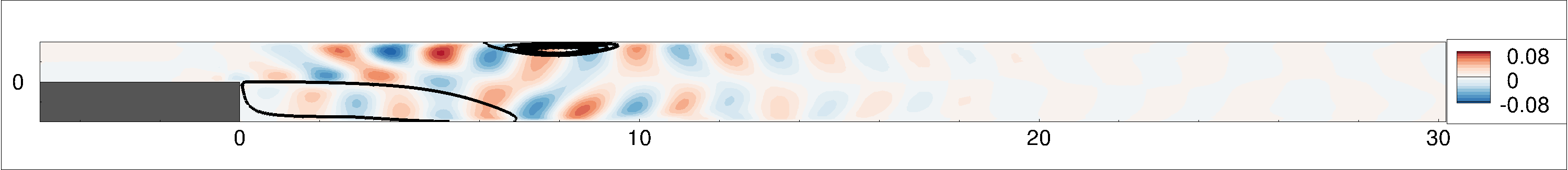}
    
    \raisebox{0.2in}{(i)}\includegraphics[trim={1cm 1cm 1cm 1cm},clip,width=300px]{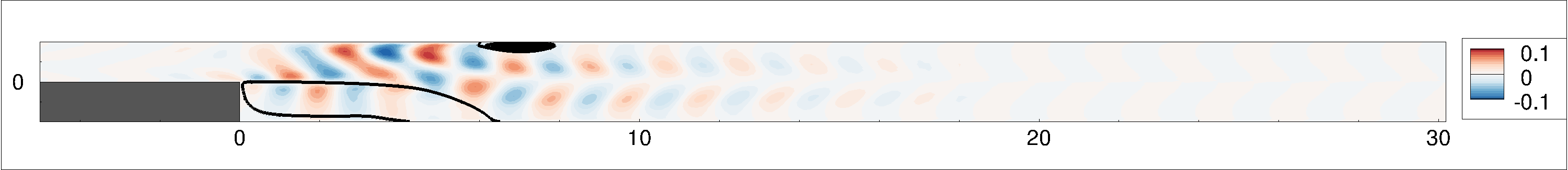}
    \raisebox{0.2in}{(j)}\includegraphics[trim={1cm 1cm 1cm 1cm},clip,width=300px]{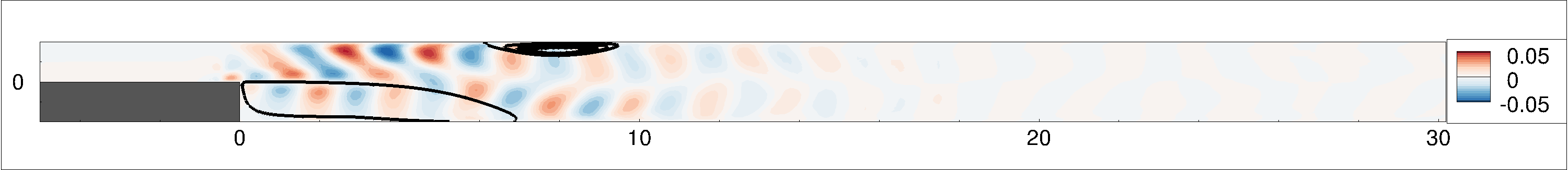}

    \raisebox{0.2in}{(k)}\includegraphics[trim={1cm 1cm 1cm 1cm},clip,width=300px]{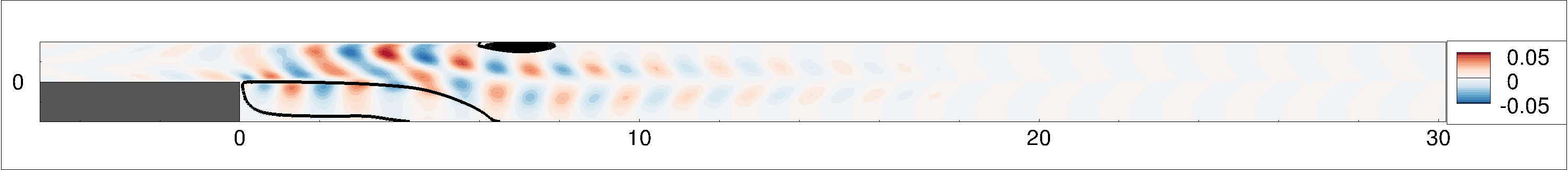}
    \raisebox{0.2in}{(l)}\includegraphics[trim={1cm 1cm 1cm 1cm},clip,width=300px]{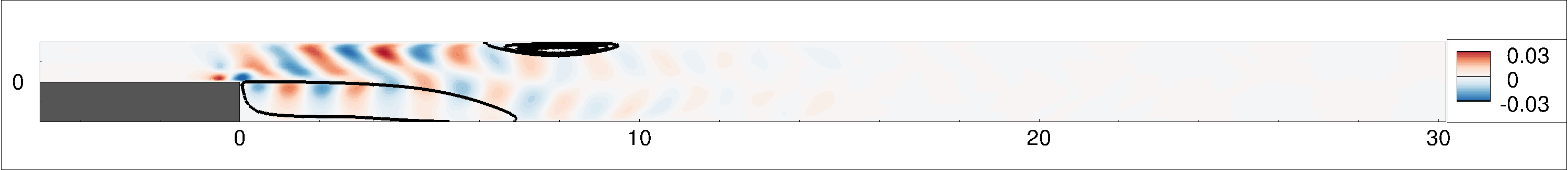}
    
	\caption{Comparison between reconstructed (a,c,e,f,i,k) and reference (b,d,f,h,j,l) fields: mean-flow (a,b) and Fourier modes at $\omega=0.2$ (c,d), $\omega=0.6$ (e,f), $\omega=1.0$ (g,h), $\omega=1.4$ (i,j),  $\omega=1.8$ (h,l).} \label{fig:unsteadyBFS-wn-resolvent}
\end{figure}

We now turn our attention to the comparison of the assimilated fields with the ones from the DNS. In figure \ref{fig:unsteadyBFS-wn-resolvent}, we compare the reconstructed mean-flow and Resolvent modes for $J^1$ and $N=5$ with the mean-flow and SPOD modes of the DNS. We can see that, although the quantitative features are not necessarily very accurate (length of upper separation bubble, amplitude of fluctuations), the qualitative features of the mean-flow and SPOD modes are well captured.
This shows that the present procedure is effective to reconstruct the overall features of the mean- and unsteady characteristics of a flow. We remind also that only very few point-wise measurements have been considered for that purpose.

\begin{figure}
	\centering
	\raisebox{0.7in}{(a)}\includegraphics[trim={0.cm 0cm 1cm 0cm},clip,height=70px]{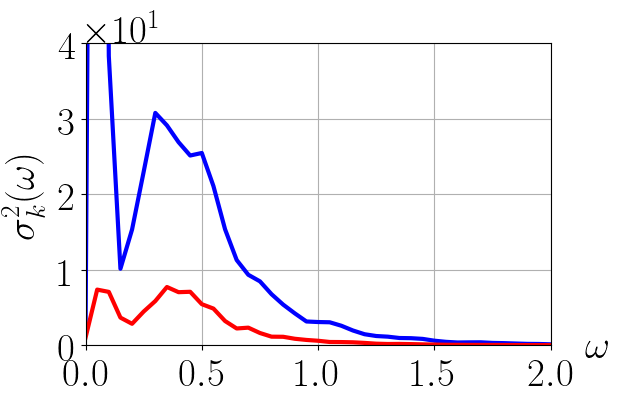}
    \raisebox{0.7in}{(b)}\includegraphics[trim={0.cm 0cm 1cm 0cm},clip,height=70px]{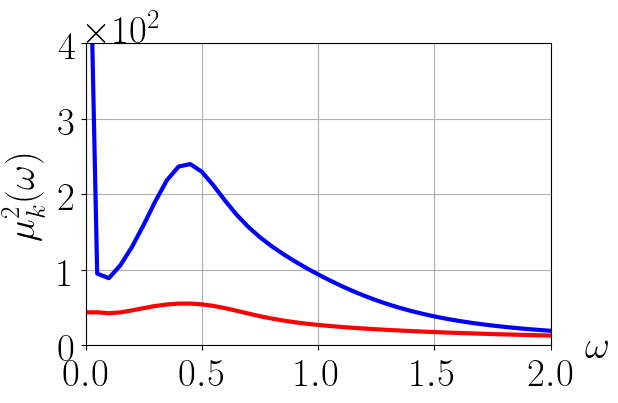}
	\raisebox{0.7in}{(c)}\includegraphics[trim={0.cm 0cm 0cm 0cm},clip,height=70px]{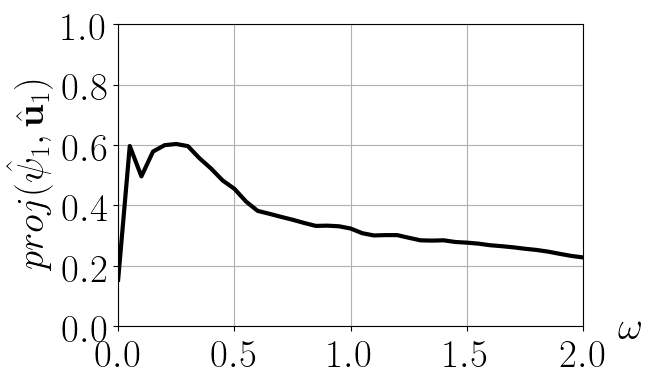}
    \caption{(a) SPOD energies $\sigma_k^2(\omega)$, (b)  Resolvent gains $\mu_k^2(\omega)$ and (c) the (normalized) projection between resolvent and SPOD leading modes as a function of frequency.}\label{fig:SPOD-resolvent-proj}
\end{figure}

A better understanding of the quality of the reconstruction can be gained by analysing the SPOD / resolvent modes and the projection coefficients between these modes.
We have plotted in figure \ref{fig:SPOD-resolvent-proj} (a) the two dominant SPOD gains as a function of frequency. We can see that a quite strong suboptimal branch exists, indicating that the dynamics is not really rank one. These suboptimal SPOD modes have not been considered in the present article, but could in principle be. Figure \ref{fig:SPOD-resolvent-proj} (b) shows that the flow exhibits a dominant resolvent mode, which is at the origin of the dominant SPOD mode. Yet, the suboptimal resolvent gain is not very weak with respect to the dominant one, which may explain why the dynamics is not strictly rank-one in terms of SPOD gains. In \ref{fig:SPOD-resolvent-proj} (c) we can see that the optimal SPOD and resolvent modes are not fully aligned, a property which is at the heart of the present reconstruction procedure. Hence, in flowfields where this alignement is even better and the dynamics closer to rank 1 (see for example turbulent jets where the alignement reaches values of 0.9 instead of 0.6 here, see \cite{schmidt2018spectral}), the reconstruction could be much better.
Note also that a more complex model with additional suboptimal modes could be considered. Also, \cite{pickering2020optimal} showed that the alignement of the optimals (and the suboptimals) could be improved in turbulent flows by considering an eddy-viscosity for the definition of the resolvent modes.

\section{Conclusions and Perspectives}

In this work, we presented a data-assimilation technique based on Resolvent modes around the mean-flow. The technique was based on a model composed of a mean-flow equation coupled with resolvent modes corresponding to left singular vectors of the Resolvent operator. The model is not closed since the amplitudes of those modes are unknown and need to be tuned with the help of measurements, which are typically takes as point-ise time-resolved probes. This technique was applied on an oscillator flow, whose frequency content was essentially monochromatic. For this reason, the data-assimilation results showed that we are indeed capable of accurately recovering not only the mean-flow but also the fluctuation around it. This technique was also applied on a more challenging noise-amplifier backward-facing step flow where the dynamics is here driven by an external white-noise forcing, leading to a broadband-frequency fluctuation. For this reason, a connection between this broadband fluctuation and our model, which is essentially peaked-frequency, needed to be established. This connection is based on the integral of the Power-Spectral Density, which is a finite quantity either for the actual reference flow and for our model. By enforcing our model to satisfy those quantities at certain point locations, where the velocity field was measured, we were able to obtain a fairly good solution that reproduced this time both mean-flow quantities and the fluctuation as well. One shortcoming of this approach was the dependency of the assimilated solution on the spatial location of the measurements. This comes from a mismatch between the SPOD modes and the Resolvent ones. Indeed, if one of those modes is stronger/weaker than the other in a specific region of the flow, we may be over/under predicting its energy, leading to erroneous results. Another shortcoming of this approach is that only the leading mode is taken into account in the model, leading to other possible models where several modes, for a given frequency, are considered in the model. Apart from the higher computational complexity, one difficulty of this approach is the tendency of suboptimal Resolvent and SPOD modes to differ. For those reasons, generally, we believe that adding an eddy-viscosity could improve the results since it usually improve the comparability of those modes (see \citet{morra2019relevance,pickering2020optimal}).


\bibliography{bib}

\end{document}